\documentclass[journal]{IEEEtran}

\usepackage{bbm}
\usepackage{cite}
\usepackage{amsmath}
\usepackage{amssymb}
\usepackage{amsfonts}
\usepackage{graphicx}
\usepackage{epstopdf}
\usepackage{algorithm}
\usepackage{algorithmic}
\usepackage{epsfig}
\usepackage{epstopdf}
\usepackage{subfigure}
\usepackage{url}

\usepackage{hyperref}
\hypersetup{
	colorlinks=true,
	linkcolor=blue,
	citecolor=blue,
	urlcolor=blue 
}

\usepackage{pifont}
\usepackage{psfrag}
\usepackage{times}
\usepackage{booktabs}
\usepackage[T1]{fontenc}
\usepackage{moreverb}
\usepackage{color}
\usepackage{setspace}
\usepackage{mathrsfs}
\usepackage{array}
\usepackage{txfonts}
\usepackage{makecell}
\usepackage{bm}
\newtheorem{Lemma}{Lemma}
\newtheorem{Theorem}{Theorem}

\newtheorem{Proposition}{Proposition}
\newtheorem{Proof}{Proof}

\begin{document}
\makeatletter
\def\ps@IEEEtitlepagestyle{%
	\def\@oddhead{%
		\scriptsize
		\parbox{\textwidth}{%
			\centering
			This article has been published in IEEE Transactions on Wireless Communications. This is the author's version which has not been fully edited and content may change prior to final publication. Citation information: DOI 10.1109/TWC.2025.3620886.
		}%
	}%
	\def\@evenhead{\@oddhead}
	\def\@oddfoot{}%
	\def\@evenfoot{\@oddfoot}%
}
\makeatother
	\title{Circular Holographic MIMO Beamforming for Integrated Data and Energy Multicast Systems
		\author{Qingxiao Huang,~\IEEEmembership{Graduate Student Member,~IEEE}, Yizhe Zhao,~\IEEEmembership{Member,~IEEE}, Jie Hu,~\IEEEmembership{Senior Member,~IEEE}, Kun~Yang,~\IEEEmembership{Fellow,~IEEE} and Yuguang Fang,~\IEEEmembership{Fellow,~IEEE}}
		\thanks{This work was supported in part by the Natural Science Foundation of China (NSFC) under Grant No. 62132004, No. 62201123, No. 62431002 and No. 62531008 in part by the Young Elite Scientists Sponsorship Program by CAST under Grant 2023QNRC001, in part by the Natural Science Foundation of Sichuan (NSFSC) under Grant 2024NSFSC1417, in part by Zhejiang Province Major Research and Development Plan (No. 2024C01062), in part by Jiangsu Major Project on Fundamental Research (Grant No.: BK20243059), in part by Gusu Innovation Project (Grant No.: ZXL2024360), in part by High-Tech District of Suzhou City (Grant No.: RC2025001), in part by Quzhou Government (Grant No. 2024D007, 2023D005), and in part by a grant from the Research Grants Council of the Hong Kong Special Administrative Region, China (Project No. CityU 11216324). {\it (Corresponding author:	Yizhe Zhao.)}}
		\thanks{Qingxiao Huang is with the School of Information and Communication Engineering, University of Electronic Science and Technology of China, Chengdu, 611731, China, and is also with the Hong Kong Jockey Club STEM Lab of Smart City and the Department of Computer Science, City University of Hong Kong, Hong Kong, email: qx.huang@my.cityu.edu.hk.}
		\thanks{Yizhe Zhao and Jie Hu are with the School of Information and Communication Engineering, University of Electronic Science and Technology of China, Chengdu, 611731, China, email: yzzhao@uestc.edu.cn and hujie@uestc.edu.cn. }
		\thanks{Kun Yang is with the State Key Laboratory of Novel Software Technology, Nanjing University, Nanjing, 210008, China, Institute of Intelligent Networks and Communications (NINE), Nanjing University (Suzhou Campus), Suzhou, 215163, China, and School of Information and Communication Engineering, University of Electronic Science and Technology of China, Chengdu, China, email: kunyang@nju.edu.cn.}
		\thanks{Yuguang Fang is with the Hong Kong Jockey Club STEM Lab of Smart City and the Department of Computer Science, City University of Hong Kong, Hong Kong, e-mail: my.Fang@cityu.edu.hk.}
	}
	\maketitle
	\begin{abstract}
		Due to the innovative application of metamaterials, holographic multiple-input multiple-output (H-MIMO) is expected to achieve a higher spatial diversity gain with lower hardware complexity. Together with the aid of a circular antenna arrangement in H-MIMO, integrated data and energy multicast (IDEM) can fully exploit the near-field channel to realize wider range of energy focusing and higher achievable rate. In this paper, we focus on beamforming design and investigate the IDEM systems that maximize the minimum rate of data users (DUs) while meeting the energy harvesting requirements for energy users (EUs). Specifically, we first derive the closed-form near-field resolution function in 3D space and show the asymptotic spatial orthogonality of near-field channel for circular antenna arrays. Then, we design an asymptotically optimal fully-digital beamformer based on the spatial orthogonality. After that, we apply the alternating optimization to develop H-MIMO beamforming scheme, where the digital beamformer is given in closed form while the analog beamformers of three different control modes are obtained numerically, respectively. Scaling schemes are also investigated to further improve the IDEM performance. Numerical results verify the correctness of the resolution function and asymptotic orthogonality and demonstrate that  the proposed beamforming schemes outperform benchmark schemes, with very low complexity.
	\end{abstract}
	\begin{IEEEkeywords}
		Integrated data and energy multicast (IDEM), holographic MIMO, circular antenna array, spatial orthogonality.
	\end{IEEEkeywords} 
	\section{Introduction}
\subsection{Background}
In the Internet of Things (IoT), seamless and intelligent connections between individuals and devices are becoming increasingly prevalent. The integration of millions, or even billions, of IoT devices with compact hardware designs, low complexity, and minimal power consumption into smart cities is driving the development of innovative applications such as smart transportation, virtual reality, and edge computing \cite{10122600,10449899}. However, this rapid progress introduces significant challenges related to meeting the energy demands of these devices. The frequent need to replace or recharge batteries not only incurs high maintenance costs but also raises safety concerns. In the context of ambient IoT, low-power devices are designed to harvest energy from environmental sources, including light, motion, heat, and, most notably, radio frequency (RF) signals \cite{RP-234058}. Utilizing existing infrastructure such as cellular base stations, television towers, and Wi-Fi access points, RF-based wireless energy transfer (WET) offers superior control and stability compared to other energy-harvesting methods \cite{8421584}. When combined with wireless data transfer (WDT), the integrated data and energy transfer (IDET) framework presents a forward-looking solution, enabling low-power IoT devices to achieve both energy self-sustainability and efficient data communication \cite{9261955}. 

In IDEM systems, to mitigate significant path loss, particularly in the WET component, massive MIMO is commonly employed to achieve substantial spatial gain. By utilizing beamforming techniques, the transmitter can precisely align the beams while enabling resource allocation in both spatial and power domains \cite{10559446}. Additionally, with advancements in metamaterials, holographic MIMO (H-MIMO) with leaky-wave antenna architectures has gained prominence\cite{PhysRevApplied.8.054048}. H-MIMO enables large-scale antenna arrays at low cost and offers higher energy efficiency due to its serial feeding structure. Moreover, H-MIMO allows for sub-wavelength antenna spacing, which facilitates near-continuous antenna apertures and enhances electromagnetic control capabilities \cite{9136592}.

With the increasing frequency bands of communication systems and the expansion of antenna array sizes, users are more likely to fall within the near-field range, especially energy users (EUs) who are typically situated closer to a transmitter \cite{10721321}. In the near-field region, electromagnetic waves can no longer be approximated as plane waves; instead, a spherical wave model must be employed, enabling near-field beam focusing \cite{10558818}. Additionally, the circular arrangement of antenna arrays provides a 360° near-field coverage, which holds great potential for advancing near-field technologies \cite{10480331}.

\subsection{Related Works}
Massive MIMO and beamforming for physical-layer multicast have been extensively studied over the years. Sidiropoulos \emph{et al.} \cite{1634819} explored fully digital MIMO transmitter-based multicast systems, proposing the classic semidefinite relaxation (SDR) algorithm to address two fundamental problems: minimizing transmission power and maximizing fairness rates. Liu \emph{et al.} \cite{9738456} introduced innovative algorithms for fairness rate maximization by integrating machine learning. Specifically, their approach involved user selection with the minimum rate using machine learning methods, followed by leveraging QR decomposition and continuous beamforming to design a beamforming scheme tailored to that user. Furthermore, Yue \emph{et al.} \cite{9145622} extended physical-layer multicast to IDET systems. Recognizing the high cost of implementing fully digital transmitters for massive MIMO in high-frequency bands, they investigated hybrid transmitter-assisted IDEM systems and developed efficient beamforming schemes to enhance both WDT and WET performance. Zhai \emph{et al.} \cite{9582739} further examined intelligent reflecting surface-assisted IDEM systems, devloping a game-theoretic approach to solve the WDT problem through SDR and the WET problem using an alternating iterative method. Zhang \emph{et al.} \cite{10453453} carried out the performance analysis of this IDEM scenario, employing stochastic geometry to evaluate energy alignment for IoT devices as well as the uplink outage probability and average rate.

Compared with fully digital and hybrid transmitters, using low-cost H-MIMO to implement ultra-large-scale MIMO arrays is more practical,  because the metamaterial elements of H-MIMO must adhere to the Lorentzian frequency response, leading to diverse hardware designs \cite{PhysRevApplied.8.054048}. Specifically, Deng \emph{et al.} \cite{10163760} introduced a hardware design for H-MIMO that supports amplitude-controlled beamforming, implementing a prototype capable of real-time video transmissions. Building on the amplitude-controlled hardware structure, they further explored the applications of H-MIMO in integrated sensing and communication (ISAC) systems \cite{9724245}, target detection and positioning systems \cite{10745216}, and low-earth-orbit satellite systems \cite{9848831}. In contrast to amplitude control, the phase-only-controlled H-MIMO is governed by a Lorentzian-constrained phase model, with the control region represented by a circle in the upper complex plane passing through the origin, delivering comparable performance gains. \cite{9738442}. Additionally, Zhang \emph{et al.} investigated H-MIMO communication systems with wireless feeding, where the size of H-MIMO was optimized to maximize the system's energy efficiency \cite{10287263}. Furthermore, An \emph{et al.} \cite{10158690} extended this architecture to multiple stacked planar arrays. This innovative stacked intelligent metasurface (SIM) architecture enables transmit precoding and receiver combining directly in the native electromagnetic wave domain, eliminating the need for numerous RF chains. Moreover, Papazafeiropoulos \emph{et al.} \cite{10534211} studied a SIM-assisted holographic communication systems, where SIMs deployed at both transceivers enable efficient wave-based precoding and combining, and proposed a fast-converging optimization scheme, achieving higher spectral efficiency compared to conventional MIMO systems. Furthermore, Li et al. \cite{10535263} applied the SIM to cell-free networks, where SIMs at distributed access points achieve efficient distributed beamforming and local detection, and proposed low-complexity scheme that outperforms conventional single-layer H-MIMO in achievable rate. Besides, Ji \emph{et al.} \cite{10628002} investigated 3D holographic MIMO communication systems, where an electromagnetic hybrid beamforming scheme was designed to achieve superdirective gain and a relatively flat beamforming gain across the spatial domain by considering array radiation patterns and coupling effects.

Ultra-large-scale H-MIMO arrays have facilitated the transition from far-field to near-field communication. Zhang \emph{et al.} \cite{9738442} analyzed near-field communication systems employing fully digital, hybrid, and H-MIMO transmitters, where beamforming enables beam focusing to enhance achievable rates in the near field. Besides, Wei \emph{et al.} \cite{10103817} investigated tri-polarized antenna-based H-MIMO systems, designing beamforming schemes based on a near-field channel model to mitigate user interference and analyzing the channel capacity of tri-polarized systems. Moreover, Wu \emph{et al.} \cite{10086987} explored H-MIMO-based WET systems, proposing multi-point beam focusing schemes validated through simulations and experimental implementations. Additionally, Wu \emph{et al.} \cite{10243590} examined communication systems with circular antenna arrays, demonstrating through performance analysis that circular arrays provide broader benefits to users at various angles in near-field communications. They also developed a near-field codebook tailored for circular arrays. Furthermore, Wang \emph{et al.} \cite{10349918} studied cylindrical H-MIMO multi-beam hardware systems, where a multi-subregion H-MIMO design was evaluated, fabricated, and experimentally tested.

Due to the large number of controllable elements in H-MIMO, it is essential to design low-complexity algorithms to reduce the computational power consumption and latency in practical systems. Specifically, Ayach \emph{et al.} \cite{6717211} studied millimeter-wave communication systems, where a hybrid beamforming design based on channel sparsity was proposed to improve achievable rates. Additionally, Yue \emph{et al.} \cite{9145622} designed low-complexity hybrid beamforming based on spatial orthogonality for IDET systems. Furthermore, Wu \emph{et al.} \cite{10123941} leveraged the spatial orthogonality of near-field channels to design a location-based multiple access system.

For holographic IDET systems, Huang \emph{et al.} \cite{10721321} investigated an ideal continuous aperture H-MIMO-assisted IDET system, where beamforming schemes based on block coordinate descent and Fourier transforms were proposed to maximize throughput while satisfying energy harvesting requirements. They also extended the lower bound of the traditional near-field range by utilizing electromagnetic channel. Furthermore, Liu \emph{et al.} \cite{10680303} explored a holographic IDET system based on anisotropy-metasurface field synthesis, where a polarization diversity scheme was designed to enhance the performance of multi-user IDET. In addition, Huang \emph{et al.} \cite{10415645} studied a Lorentzian-phase constrained H-MIMO IDET system, where an iterative weighted minimum mean-square error-based algorithm was developed to maximize throughput while ensuring the energy harvesting requirements.

\subsection{Motivation and Contributions}
However, the existing researches have the following drawbacks.
\begin{itemize}
	\item Although electromagnetic theory-based studies have demonstrated the superiority of H-MIMO for IDET, the assumption of nearly continuous apertures is not practical. This requires a detailed and refined design of a practical H-MIMO hardware architecture.
	\item Near-field H-MIMO typically involves a large number of antennas. Traditional algorithms, usually based on successive convex approximation, manifold optimization, or other approaches, are highly complex and unsuitable for ultra-large MIMO arrays.
	\item While performance analysis and multiple access design of near-field channel orthogonality have been conducted for planar arrays, further research is required to apply the orthogonality to the complex holographic beamforming designs. Additionally, the general characterization of channel orthogonality in three-dimensional space for circular arrays still requires further investigation. 
\end{itemize}
\begin{figure*}[t]
	\centering
	\includegraphics[width=140mm]{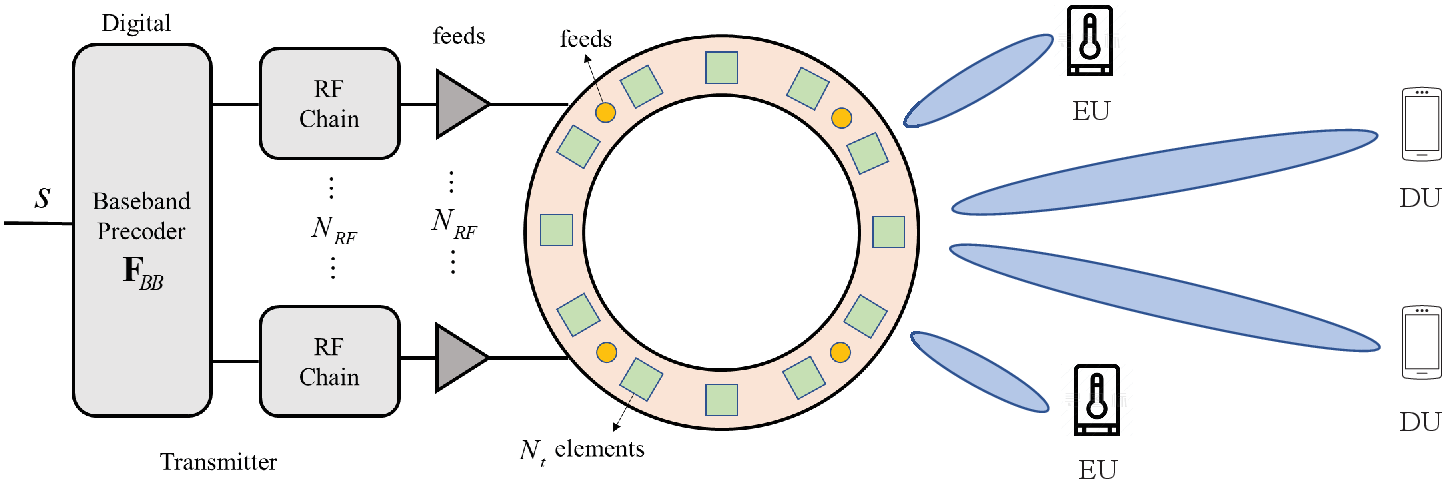}
	\caption{System model.}
	\label{fig_system}
\end{figure*}
In light of these, our novel contributions are summarized as follows:
\begin{itemize}
	\item We investigate a H-MIMO-assisted IDET multicast system. By designing beamforming, we maximize the minimum rate of information users while satisfying the transmit power constraint and the minimum energy harvesting requirement for energy users, thereby improving fairness achievable rate.
	\item Under a circular antenna array, we derive a closed-form expression for the resolution function between any two points in three-dimensional space. Furthermore, through an upper bound and asymptotic analysis, we demonstrate the spatial asymptotic orthogonality of near-field channels.
	\item Utilizing spatial orthogonality, the low-complexity beamforming schemes are designed  for three different control modes in H-MIMO. We also further improve these schemes by using demonstrated equivalent scaling of the analog domain constraints.
	\item Simulation results confirm the validity of the derived resolution function and the asymptotic orthogonality. Moreover, circular antenna arrays offer a wider near-field range in the angular domain compared to linear arrays. Our approach also enables multi-point near-field focusing. 
\end{itemize}

The rest of the paper is organized as follows: In Section II, we investigate the system modeling and problem formulation of the circular H-MIMO aided IDEM system, while the near-field spatial orthogonality for circular array is explored in Section III. The beamforming design based on spatial orthogonality is proposed in Section IV. Then, our numerical results are presented in Section V, followed by the conclusions in Section VI.

Notations: $(\cdot)^H$ denotes conjugate transpose operations; $\mathbb{E} \{ \cdot \}$ represents the statistical expectation; $|a|$ and $||\mathbf{a}||$ are the magnitude and norm of a scalar $a$ and vector $\mathbf{a}$, respectively; $||\mathbf{A}||$ denotes the Frobenius norm of the matrix $\mathbf{A}$; $\mathbf{A}(i,j)$ represents the specific element in the $i$-th row and $j$-th column of $\mathbf{A}$; $\mathbf{a}(i)$ denotes the $i$-th element of the vector $\mathbf{a}$; $\mathbf{A}(i,:)$ represents the row vector formed by the elements of the $i$-th row of matrix $\mathbf{A}$; ``:='' is signified as ``is defined as''; and $ \mathfrak{R}(\cdot) $ and $ \mathfrak{I}(\cdot) $ represent the real and imaginary parts of their arguments, respectively.

\section{Circualr H-MIMO assisted Integrated Data and Energy Multicast System}
\subsection{System Model}

As shown in Fig. \ref{fig_system}, a downlink IDET multicast system is investigated, where the transmitter serves $ K $ data users (DUs) and $ L $ energy users (EUs), which are both equipped with single antenna. Let $ \mathcal{D} $ and $ \mathcal{E} $ represent the set of DUs and EUs, respectively. Then, the set of all users is given by $ \mathcal{U}=\mathcal{D} \cup \mathcal{E} $. The transmitter multicasts common messages to all users. It is assumed that the transmitted message $ s $ follows a complex Gaussian distribution, \textit{i.e.}, $ s \sim \mathcal{CN}(0,1) $.

Then, the received signals can be expressed as
\begin{align}\label{eq:received signal}
	y_k &=\sqrt{P_t} \mathbf{h}^H_k\mathbf{Q}\mathbf{P}\mathbf{b}s+n_k 
\end{align}
where $ P_t $ is the transmit power, $ \mathbf{b}\in \mathbb {C}^{N_{RF}\times1} $ is the baseband digital beamfomer, $ \mathbf{Q}\in \mathbb {C}^{N\times N} $ is the analog beamfomer of H-MIMO, $ \mathbf{P} \in \mathbb {C}^{N\times N_{RF}}$ is the propagation matrix with respect to the propagation in the waveguide from the feeds to the antennas, and $ N_{RF} $ is the number of radio frequency (RF) chains.

Specifically, $ \mathbf{Q}=\mathrm{diag}(\mathbf{q})=\mathrm{diag}(q_1,q_2,\dots,q_N) $, a dialog matrix, wherein every element satisfies the hardware constraints. Let the set of possible values of $ q_n $ as $ \mathcal{Q} $. There are normally three types of control modes \footnote{The three control modes are all based on the same physical framework. A propagating reference field within the waveguide establishes a global linear phase gradient across the aperture, while each subwavelength radiating element can be modeled as a single-resonance polarized magnetic dipole, whose polarizability follows the Lorentzian formula.} \cite{PhysRevApplied.8.054048}:
\begin{itemize}
	\item Amplitude only, \textit{i.e.}, $ 0<q<1 $;
	\item Binary amplitude, \textit{i.e.}, $ q = 0 $ or $ 1 $;
	\item Lorentzian-constrained phase, \textit{i.e.}, $ q = \frac{j+e^{j\phi}}{2}, \phi\in[0,2\pi] $. 
\end{itemize}

The elements in propagation matrix $ {P} \in \mathbb {C}^{N\times N_{RF}}$ satisfies
\begin{align}
	{P}_{m,n} = e^{(\gamma+j\beta)l_{m,n}}
\end{align}
where $ \beta $ is the wavenumber, $ \gamma $ is the waveguide attenuation coefficient, $ l_{m,n} $ is the path of the electromagnetic wave clockwise along the circle waveguide, from the $ n $-th feed to the $ m $-th antenna.

Based on the received signal of $ DU_k $ expressed in Eq.(\ref{eq:received signal}), SNR of $ DU_k $ is expressed as
\begin{equation} \label{eq:5}
	\gamma_k=\frac{P_t||\mathbf{h}^H_k\mathbf{Q}\mathbf{P}\mathbf{b}||^2}{\sigma^2}, \ \ k \in \mathcal{K}.
\end{equation}
and the achievable rate of $ DU_k $ is expressed as
\begin{equation} \label{eq:6}
	R_k=\log_2(1+\gamma_k),\ \ k \in \mathcal{K}.
\end{equation}
The received energy harvesting of $ EU_l $ is expressed as
\begin{equation} \label{eq:7}
	E_l=||y_l||^2=P_t||\mathbf{h}^H_k\mathbf{Q}\mathbf{P}\mathbf{b}||^2,\ \ l \in \mathcal{E}.
\end{equation}

\subsection{Problem Formulation}
In order to ensure fairness among all the users, the achievable rate of the system is defined as the minimum achievable rate among DUs. Our goal is to maximize the achievable rate of the system, while satisfying individual wireless charging requirement of EUs. This problem can be formulated as
\begin{align}
	\text{ (P1):} \quad  &R_{\mathrm{min}}\triangleq \max_{ \substack{\mathbf{Q}, \mathbf{b}} } \ { \min \limits_{1\leq k \leq K}} R_k, \label{eq:object1} \\
	\text{s.t.} \  &
	\left\| \mathbf{Q}\mathbf{P}\mathbf{b}\right\| ^2= 1,  \tag{\ref{eq:object1}a} \label{eq:object1-a} \\
	& q_n\in \mathcal{Q}, \forall n, \tag{\ref{eq:object1}b} \label{eq:object1-b}\\
	& {E}_l\geq E_0, l \in \mathcal{E}  \tag{\ref{eq:object1}c} \label{eq:object1-c}
\end{align}
where constraint (\ref{eq:object1}a) ensures that the transmitter operates within the maximum transmission power limit, while constraint (\ref{eq:object1}b) defines the analog domain restrictions under various control methods. Moreover, constraint (\ref{eq:object1}c) guarantees that the energy harvested by EUs is no less than the minimum energy requirement \(E_0\). It can be observed that (P1) is a non-convex optimization problem, since the constraints (\ref{eq:object1}b) and  (\ref{eq:object1}c) are non-convex. Moreover, the variables are coupled together and the objective function is intractable. Besides, the analog domain constraints vary across different control methods and require joint design in the digital domain, which poses a significant challenge. Furthermore, considering the extremely large number of antennas involved, brute-force approaches such as exhaustive search over discretized continuous variables become computationally infeasible due to their exponentially growing complexity. Therefore, effective relaxation and reformulation are required to tackle the above non-convex optimization problem.
 
\section{Near-field Spatial Orthogonality for Circular Antenna Array}
	\begin{figure}[t]
	\centering
	\includegraphics[width=90mm]{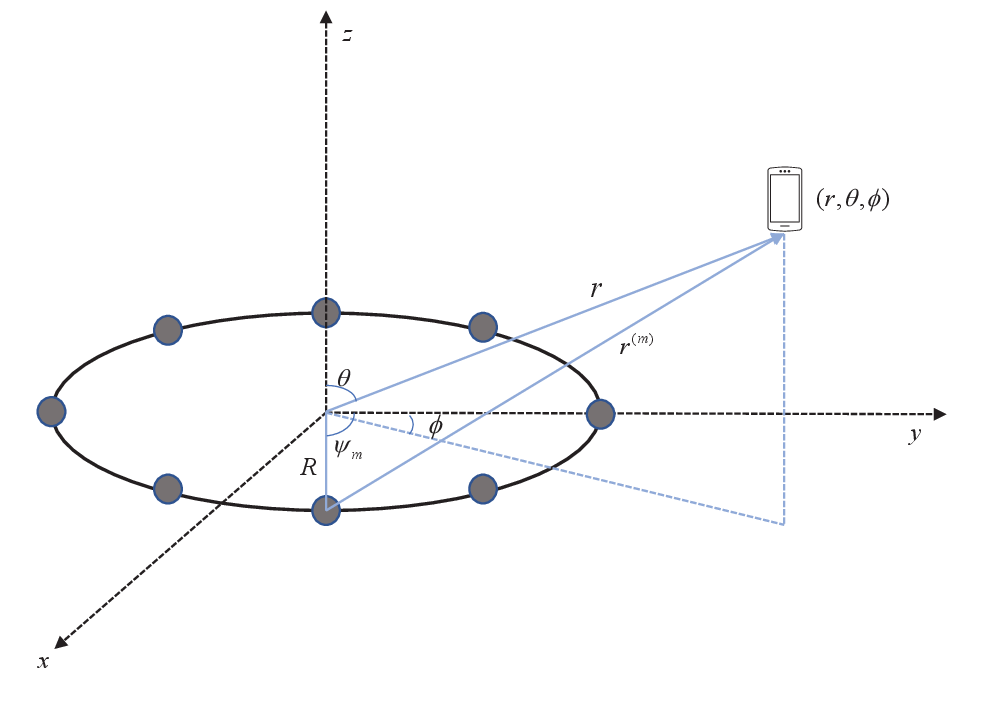}
	\caption{The geometric relationship between the circular antenna array and the user in three-dimensional space.}
	\label{fig_channel}
\end{figure}
	In the three-dimensional space depicted in Fig. \ref{fig_channel}, the circular antenna array is centered at the origin of either the Cartesian or spherical coordinate system, with its plane aligned parallel to the horizontal plane. Consequently, a user's position in the spherical coordinate system is denoted as \((r, \theta, \phi)\). Assuming that a user is equipped with a single antenna, the near-field channel can be expressed as\cite{10243590}:
	\begin{equation}\label{eq:channel}
		\mathbf{h}=\sqrt{N}\alpha^{[0]} \mathbf{a}_t(r^{[0]},\theta^{[0]},\phi^{[0]})+\sqrt{\frac{N}{I}}\sum_{i=1}^{I}\alpha^{[i]} \mathbf{a}_t(r^{[i]},\theta^{[i]},\phi^{[i]})
	\end{equation}
	where the first term represents the Line-of-Sight (LoS) path, while the second term represents the Non-Line-of-Sight (NLoS) paths. \(I\) and $ N $ represent the number of NLoS paths and transmit antennas, respectively. Additionally, \(\alpha^{[i]}\) denotes the complex path gain of the \(i\)-th path, while
	 \(\mathbf{a}_{t}(r, \theta, \phi)\) represents the channel response vector in polar domain, specifically expressed as
	\begin{align}
\mathbf{a}_t(r,\theta,\phi)=\frac{1}{\sqrt{N}}[e^{j\frac{2\pi}{\lambda}(r-r^{(1)})},\dots,e^{j\frac{2\pi}{\lambda}(r-r^{(N)})}]^T
	\end{align}
	where $ r^{(n)} $ is the distance between the user and the $ n $-th antenna, specifically expressed as
	
\begin{align} \label{eq:3}
r^{(n)}&=\sqrt{R^2+r^2\sin^2\theta-2rR\sin\theta\cos(\phi-\psi_n)+(r\cos\theta)^2}\nonumber\\
&=\sqrt{r^2+R^2-2rR\sin\theta\cos(\phi-\psi_n)} \nonumber\\
&\overset{\text{(a)}}{\approx} r-R\sin(\theta)\cos(\phi-\psi_n)+\frac{R^2}{2r}\left( 1-\cos^2(\phi-\psi_n)\sin^2\theta\right) \nonumber\\
&-\frac{R^3}{2r^2}\sin\theta\cos(\phi-\psi_n)-\frac{R^4}{8r^3}
\end{align}
where (a) is obtained by utilizing $ \sqrt{1+x}=1+x/2-x^2/8+\mathcal{O}(x^2) $.
 
Channel estimation for H-MIMO can be performed by exploiting the rank deficiency of the spatial correlation matrix, which is induced by the array geometry, to facilitate subspace-based denoising and improve estimation accuracy. Alternatively, the inherent sparsity of H-MIMO channels in the angular, delay, or polar domains can be leveraged through compressive sensing and parametric modeling to significantly reduce pilot overhead \cite{10130641}.

To evaluate the spatial orthogonality between the channel response vectors of two users in three-dimensional space, the resolution function can be used, which is defined as 
\begin{align}\label{eq:def_resolution}
	&\Delta(r_1,\theta_1,\phi_1,r_2,\theta_2,\phi_2)=\left| \mathbf{a}_t^H(r_1,\theta_1,\phi_1)\mathbf{a}_t(r_2,\theta_2,\phi_2)\right|  \nonumber \\
	&=\left|\frac{1}{N}\sum_{n=1}^{N}e^{j\frac{2\pi}{\lambda}\left( \sqrt{r_1^2+R^2-2r_1R\sin\theta_1\cos(\phi_1-\psi_n)}-r_1-\sqrt{r_2^2+R^2-2 r_2 R\sin\theta_2\cos(\phi_2-\psi_n)}+r_2 \right) } \right| 
\end{align}
Typically, the smaller the function value, the better the asymptotic orthogonality. To facilitate the subsequent analysis, we start with deriving the closed-form expression of the resolution function in  {\it Lemma 1}.

\begin{Lemma} \textbf{(Near-field resolution function for circular antenna array):}
	Considering two coordinates $ (r_1,\theta_1,\phi_1)$ and $(r_2,\theta_2,\phi_2) $ in the polar domain, the resolution function between them can be expressed as
	\begin{equation}\label{eq:resolution function}
		\Delta(r_1,\theta_1,\phi_1,r_2,\theta_2,\phi_2)=\left| \xi_5\sum_{n=-\infty}^{\infty}j^n J_n(\xi_3)J_{2n}(\xi_1)e^{jn(2\xi_2-\xi_4-\pi/2)}\right| 
	\end{equation}
where $ \xi_1:=\sqrt{\eta_1^2+\eta_2^2} $, $ \xi_2:= \arctan\frac{\eta_1}{\eta_2}$, $ \xi_3:=\sqrt{\eta_3^2+\eta_4^2} $,
\begin{align}
	&\xi_4:= \arctan\frac{\eta_3}{\eta_4},  \xi_5:=e^{j\frac{\pi R^2}{\lambda}\left(\frac{\sin^2\theta_2}{2r_2}-\frac{\sin^2\theta_1}{2r_1}+\frac{1}{r_1}-\frac{1}{r_2}\right)},\nonumber\\
	&\eta_1:= \frac{2\pi R}{\lambda} \left( \sin\theta_2\cos\phi_2-\sin\theta_1\cos\phi_1\right),\nonumber\\
	&\eta_2:=\frac{2\pi R}{\lambda}\left( \sin\theta_2\sin\phi_2-\sin\theta_1\sin\phi_1\right), \nonumber\\
	&\eta_3:= \frac{\pi R^2}{2\lambda}\left( \frac{\sin^2\theta_2}{r_2}\cos(2\phi_2)-\frac{\sin^2\theta_1}{r_1}\cos(2\phi_1)\right), \text{and} \nonumber\\
	&\eta_4:=  \frac{\pi R^2}{2\lambda}\left( \frac{\sin^2\theta_2}{r_2}\sin(2\phi_2)-\frac{\sin^2\theta_1}{r_1}\sin(2\phi_1)\right).\nonumber
\end{align}
\end{Lemma}
\begin{IEEEproof}
Please refer to Appendix A.
\end{IEEEproof}

With the closed-form expression of the resolution function obtained, it can be shown, as stated in {\it Lemma 2}, that the channel response vectors exhibit spatial asymptotic orthogonality as the number of antennas increases.

\begin{Lemma}
	\textbf{(Near-field asymptotic spatial orthogonality for channel response vectors of circular antenna array):} 
	\begin{equation}
			\lim_{{N \rightarrow \infty}} \mathbf{a}_t^H(r_1,\theta_1,\phi_1)\mathbf{a}_t(r_2,\theta_2,\phi_2)=0.
	\end{equation}
\end{Lemma}

\begin{IEEEproof}
	Please refer to Appendix B.
\end{IEEEproof}

After demonstrating the orthogonality of the channel response vectors, the asymptotic spatial orthogonality of the near-field channel for circular antenna array can be subsequently established in Theorem 1.
\begin{Theorem}
	\textbf{(Asymptotic spatial orthogonality of near-field channel for circular antenna array):} 
	\begin{equation}
		\lim_{\substack{N \rightarrow \infty}}\frac{1}{N}\mathbf{h_1}^H\mathbf{h_2}=0.
	\end{equation}
\end{Theorem}
\begin{IEEEproof}
	Please refer to Appendix C.
\end{IEEEproof}
So far, we have shown that asymptotic spatial orthogonality can be maintained in near-field channels and under the configuration of circular arrays. In strong LoS path scenarios, such as radar sensing and ultra-high-frequency communications, channel orthogonality can be approximated using the closed-form expression of the resolution function derived earlier. In the next section, we capitalize on this spatial orthogonality to develop low-complexity and efficient beamforming schemes for IDET systems.

\section{Circular Beamforming Design based on Asymptotic Spatial Orthogonality}
\subsection{Asymptotically optimal transmitter design.}
Before solving (P1), we first consider the beamforming design for the asymptotically optimal fully-digital transmitter \( \mathbf{f} \) with a circular antenna array. For the fully-digital transmitter, the actual hardware constraints of H-MIMO can be ignored, and we have \( \mathbf{f} = \mathbf{Q} \mathbf{P} \mathbf{b} \). Thus, the problem can be formulated as:
\begin{align}
	\text{ (P2):} \quad  & \max_{ \substack{ \mathbf{f}} } \ { \min \limits_{1\leq k \leq K}} R_k=\log_2(1+\frac{P_t}{\sigma^2}||\mathbf{h}^H_k\mathbf{f}||^2), \label{eq:object2} \\
	\text{s.t.} \  &
	||\mathbf{f}||^2= 1,  \tag{\ref{eq:object2}a} \label{eq:object2-a} \\
	& {E}_l\geq E_0, l \in \mathcal{E}  \tag{\ref{eq:object2}b} \label{eq:object2-b}
\end{align}
\begin{Proposition}
	The asymptotically optimal beamformer of the  full-digital transmitter can be expressed as 
	\begin{equation}\label{eq:21}
		\mathbf{f} = \sum_{k=1}^{K}\omega_k \mathbf{h}_k + \sum_{l=K+1}^{K+L}\omega_l \mathbf{h}_l, k \in \mathcal{D}, l \in \mathcal{E}.
	\end{equation} 
	where 
	\begin{align}
		\omega_k &= \sqrt{\frac{1-\sum_{l=K+1}^{K+L}\omega_l^2 \mathbf{h}_l^H\mathbf{h}_l}{(\mathbf{h}^H_k\mathbf{h}_k)^2\sum_{k=1}^{K}\frac{1}{\mathbf{h}^H_k\mathbf{h}_k}}},\label{eq:22} \\
		\omega_l &= \sqrt{\frac{E_0}{P_t(\mathbf{h}^H_{l}\mathbf{h}_{l})^2}}.\label{eq:23}
	\end{align}
\end{Proposition}
\begin{IEEEproof}
Please refer to Appendix D.
\end{IEEEproof}
\subsection{Circular H-MIMO beamforming design}
Although fully-digital transmitters offer the highest degree of control and a straightforward solution, their use in high-frequency and near-field scenarios with large antenna arrays is impractical due to the prohibitively high hardware costs. Fortunately, H-MIMO with a hybrid analog-digital architecture can enable the cost-effective implementation of massive antenna arrays to support near-field transmissions. When a fully-digital beamformer $ \mathbf{f} $ is given, the H-MIMO counterpart can be designed by minimizing the Euclidean distance, which can be expressed as
\begin{align}
	\text{ (P3):} \quad & O_{e} \triangleq \min_{ \substack{\mathbf{Q}, \mathbf{b}} } \  \left\| \mathbf{f}-\mathbf{Q}\mathbf{P}\mathbf{b} \right\| , \label{eq:object3} \\
	\text{s.t.} \  & (\ref{eq:object1}a), (\ref{eq:object1}b).  \nonumber
\end{align}

The alternating optimization is conceived to solve problem (P3), where the digital and analog  beamfomer of H-MIMO are separated and optimized iteratively by varying one while fixing the other as the constant.
\subsubsection{Digital beamformer desigen}
By fixing the analog beamformer of H-MIMO $ \mathbf{Q} $, the optimal solution to the digital beamformer $ \mathbf{b} $ can be obtained by 
\begin{equation}\label{eq:29}
	\mathbf{b}=\arg\min_{ \substack{ \mathbf{b}} } \ \left\| \mathbf{f}-\mathbf{Q}\mathbf{P}\mathbf{b} \right\|=(\mathbf{Q}\mathbf{P})^{-1}\mathbf{f},
\end{equation}
which is the least-square solution.
\subsubsection{Analog beamformer desigen}
By fixing the digital beamformer of H-MIMO $ \mathbf{b} $, the subproblem of optimizing the analog beamformer $ \mathbf{Q} $ can be reformulated as
\begin{align}
	\text{ (P4):}  & \min_{ \substack{\mathbf{q}} }\sum_{i=1}^{N} \ \left\| \mathbf{f}(i)- \mathbf{P}(i,:)\mathbf{b}\mathbf{q}(i) \right\| , \label{eq:object4} \\
	\text{s.t.} \  & \mathbf{q}(i)\in \mathcal{Q}, \forall n, \tag{\ref{eq:object4}b}  \nonumber
\end{align}
For the simplicity, let $ \mathbf{x}(i)=\frac{\mathbf{f}(i)}{\mathbf{P}(i,:)\mathbf{b}} $, the corresponding beamforming schemes for three different control modes of H-MIMO analog beamformer can be experessed as follows: 
\begin{itemize}
	\item Amplitude only:
	\begin{align}
		\mathbf{q}(i)=\left\{
		\begin{array}{rcl}
			1,       &      & {\Re(\mathbf{x}(i))      >      1,}\\
			\Re(\mathbf{x}(i)),    &      & {0 < \Re(\mathbf{x}(i)) \leq 1,}\\
			0,     &      & {\Re(\mathbf{x}(i)) \leq 0 .}
		\end{array} \right.
	\end{align}
	\item Binary amplitude:
\begin{align}
	\mathbf{q}(i)=\left\{
	\begin{array}{rcl}
		1,       &      & {\Re(\mathbf{x}(i))      >      1/2,}\\
		0,     &      & {\Re(\mathbf{x}(i)) \leq 1/2. }
	\end{array} \right.
\end{align}
	\item Lorentzian-constrained phase:
	\begin{equation}
		\mathbf{q}(i)=\frac{j+e^{j\phi(i)}}{2}, 
	\end{equation}
\end{itemize}
where $ \phi(i)=\arg(\mathbf{x}(i)-\frac{1}{2}j) $. Then, the power constraint (\ref{eq:object1}a) can be satisfied by the normalization operation $ \mathbf{b}= \frac{1}{\left\| \mathbf{Q}\mathbf{P}\mathbf{b} \right\| }\mathbf{b}$. However, the performance of the Euclidean distance approximation is limited by the strong constraints of the analog beamformer, for which we propose the following relaxated constraint scheme.
\begin{Proposition}
	The hardware constraints for H-MIMO in the analog domain can be equivalently scaled as 
	 \begin{itemize}
		\item Amplitude only, \textit{i.e.}, $ 0<q<a $;
		\item Binary amplitude, \textit{i.e.}, $ q = 0 $ or $ a $;
		\item Lorentzian-constrained phase, \textit{i.e.}, $ q = \frac{a(j+e^{j\phi})}{2}, \phi\in[0,2\pi] $. 
	\end{itemize}
where $ a \in \mathbb {R}^{+} $ is the scaling parameter.
\end{Proposition}
\begin{Proof}
	Since the transmit power constraint is $ 	\left\| \mathbf{Q}\mathbf{P}\mathbf{b}\right\| ^2= 1  $, when the analog beamformer is scaled as $ \mathbf{Q}^{'} = a\mathbf{Q} $, the digital beamformer can be scaled as $ \mathbf{b}^{'}=\frac{1}{a}\mathbf{b} $ to satisfy the power constraint. Furthermore, the scaling of the digital beamformer can be achieved automatically by the normalization operation $ \mathbf{b}= \frac{1}{\left\| \mathbf{Q}\mathbf{P}\mathbf{b} \right\| }\mathbf{b}$. After obtaining the converged relaxed solution $ \mathbf{Q}' $, $ \mathbf{b}^{'} $ and the scaling parameter \(a\) through iteration, appropriate scaling can be applied in both the analog and digital domains to satisfy the original analog domain constraints, \textit{i.e.}, $ \mathbf{Q} = \mathbf{Q}^{'}/a, \mathbf{b}=a\mathbf{b}^{'} $.
\end{Proof}

Then, the corresponding analog beamforming schemes for Amplitude only and Lorentzian-constrained phase can be rewritten as
\begin{itemize}
	\item Amplitude only:
	\begin{align}\label{34}
		\mathbf{q}(i)=\left\{
		\begin{array}{rcl}
			\Re(\mathbf{x}(i)),    &      & {\Re(\mathbf{x}(i))      >      0,}\\
			0,     &      & {\Re(\mathbf{x}(i)) \leq 0 ,}
		\end{array} \right. 
	\end{align}

	where the scaling parameter $ a  = \Re(\mathbf{x}(i)) $.
	
	\item Binary amplitude:
	
	The optimal analog beamformer with binary amplitude can be obtained by solving the following problem
	
	\begin{align}
		\text{ (P5):}  & \min_{ \substack{\mathbf{q}},a }\sum_{i=1}^{N} \ \left\| \mathbf{x}(i)- \mathbf{q}(i) \right\| , \label{eq:object5} \\
		\text{s. t.} \  & \mathbf{q}(i)=0 \ \mathrm{or} \ a, \forall i, \tag{\ref{eq:object5}b}  \nonumber
	\end{align}
	The optimal $ \mathbf{q}(i) $ can be obtained as 
	\begin{align}\label{36}
		\mathbf{q}(i)=\left\{
		\begin{array}{rcl}
			a,       &      & {\Re(\mathbf{x}(i))      >      a/2,}\\
			0,     &      & {\Re(\mathbf{x}(i)) \leq a/2. }
		\end{array} \right.
	\end{align}
Substituting (36) into (P5), we have 
\begin{align}\label{eq:37}
	a=&\arg \min_{a} \sum_{\Re(\mathbf{x}(i))\leq a/2}\left|\mathbf{x}(i) \right| +  \sum_{\Re(\mathbf{x}(i))>a/2}\left|\mathbf{x}(i)-a \right| \nonumber\\
	=&\arg \min_{a}\sum_{\Re(\mathbf{x}(i))      >      a/2}\left|\Re(\mathbf{x}(i))-a \right|. 
\end{align}
Then, the scaling parameter $ a  $ can be obtained from the median of $ \left\{  \Re(\mathbf{x}(i)) \mid \Re(\mathbf{x}(i)) > a/2 \right\}  $.
Note that the summation term in Eq. (\ref{eq:37}) satisfies the condition $ \Re(\mathbf{x}(i))> a/2 >0 $, thus the scaling parameter $ a  $ is only affected by $ \mathcal{A}= \left\{  \Re(\mathbf{x}(i)) \mid \Re(\mathbf{x}(i)) > 0 \right\} $ and the initial value of $ a $ can be obtianed from $ \mathcal{A} $. Assuming that $ \mathbf{x}(j) \in \mathcal{B}=\left\{  \mathbf{x}(i) \mid \Re(\mathbf{x}(i)) < a/2 \right\} $, then the optimal scaling parameter $ a  $ can be obtained in \textbf{Algorithm 1}: 
		\begin{algorithm}[!]
			\linespread{1}
			\caption{Obtaining the scaling parameter $ a $ for Binary amplitude circular H-MIMO}
			\small
			\begin{algorithmic}[1]
				\STATE Initialise $ \mathcal{A}= \left\{  \mathbf{x}(i) \mid \Re(\mathbf{x}(i)) > 0 \right\}, m=\left| \mathcal{A} \right|  $;
				\REPEAT
				\STATE $ a=\frac{1}{m} \sum_{i} \mathbf{x}(i) $;
				\STATE Remove $ \mathcal{B}=\left\{  \mathbf{x}(i) \mid \Re(\mathbf{x}(i)) < a/2 \right\} $ from $ \mathcal{A} $;		
				\UNTIL $ \mathcal{B} = \emptyset $
				\STATE Return $ a$. 
			\end{algorithmic}
		\end{algorithm}

	\item Lorentzian-constrained phase: 
	
	The optimal analog beamformer with Lorentzian-constrained phase can be obtained from
	\begin{equation}\label{eq:38}
		(a,\phi(i))=\arg \min_{a>0,  \phi(i)} \sum_{i=1}^{N}\left| \frac{a(j+e^{j\phi(i)})}{2}-\mathbf{x}(i) \right| .
	\end{equation}
    Then, the optimal $ \phi(i) $ can be obtained as
    \begin{equation}\label{eq:39}
    	\phi(i)=\arg(2\mathbf{x}(i)-aj).
    \end{equation} 
    Substituting (\ref{eq:39}) into (\ref{eq:38}), we have 
    \begin{equation}\label{eq:40}
    	a=\arg \min_{a>0} \sum_{i=1}^{N}\left|\sqrt{\mathbf{x}(i)^2-a\Im(\mathbf{x}(i))+\frac{a^2}{4}} -\frac{a}{2}\right| ,
    \end{equation}
    which is a single variable non-convex problem. Let $ 	\widetilde{\mathbf{p}} = \arg \max_{i,\Im(\mathbf{x}(i))>0} \left| \mathbf{x}(i)\right| $. For simplicity, the suboptimal scaling parameter $ a $ can be obtained from 
	\begin{equation}
	\frac{a(j+e^{j\phi})}{2}=\widetilde{\mathbf{p}},
	\end{equation}
Then, we have
\begin{align}\label{42}
	\left\{	
	\begin{array}{lr}
		a=\frac{\widetilde{\mathbf{p}}^2}{2\Im(\widetilde{\mathbf{p}})},          \\
		\phi=-2\arctan \frac{\Re(\widetilde{\mathbf{p}})-\Im(\widetilde{\mathbf{p}})}{\Re(\widetilde{\mathbf{p}})+\Im(\widetilde{\mathbf{p}})},            
	\end{array} \right.
\end{align}
Finally, the analog beamformer with Lorentzian-constrained phase can be expressed as
	\begin{equation}\label{43}
		\mathbf{q}(i)=\frac{a(j+e^{j\phi(i)})}{2}, 
	\end{equation}
where $ a=\frac{\widetilde{\mathbf{p}}^2}{2\Im(\widetilde{\mathbf{p}})} $ and $ \phi(i)=\arg(2\mathbf{x}(i)-aj) $. In general, genetic algorithms work well for multi-modal non-convex problems with one variable. Therefore, the genetic algorithms for Eq. (\ref{eq:40}) is used as a benchmark to demonstrate the effectiveness of the low-complexity suboptimal $ a $ in Eq. (\ref{42}).

After obtaining the analog beamformer $ \mathbf{Q}=\mathrm{diag}(\mathbf{q}) $ of H-MIMO and the scaling parameter $ a $, the digital beamformer $ \bar{\mathbf{b}}$ can be normalized as
\begin{equation}\label{44}
	\bar{\mathbf{b}}= \frac{1}{\left\| \mathbf{Q}\mathbf{P}\mathbf{b} \right\| }\mathbf{b},
\end{equation}
which keeps the power constraint satisfied. Furthermore, after convergence, the solution satisfying the analog domain hardware constraints can be obtained by
\begin{equation}\label{eq:45}
 \mathbf{Q}^{*} = \mathbf{Q}/a ,\,\,\,\,\, \mathbf{b}^{*} = a\mathbf{b} .
\end{equation}

For clarity, the details of our algorithm for obtaining the optimal digital beamfomer $ \mathbf{b}$ and analog beamfomer $ \mathbf{Q} $ of H-MIMO are summarized in \textbf{Algorithm 2}.
\begin{algorithm}[!t]
	\linespread{1}
	\caption{Circular H-MIMO Beamforming for Integrated Data and Energy Multicast System}
	\small
	\begin{algorithmic}[1]
		\REQUIRE ~~\
		The channels of users $\{ \mathbf{h}_k\mid  k \in \mathcal{U} \}$;
		\ENSURE ~~\
		The minimum achievable rate $ R_{\mathrm{min}} $ of DUs, the digital beamformer $ \mathbf{b}\in \mathbb {C}^{N_{RF}\times1} $ and analog beamformer $ \mathbf{Q}\in \mathbb {C}^{N\times N} $ of H-MIMO;
		\STATE Obtain the optimal digital beamforming  $ \mathbf{f}$ by (\ref{eq:21}), (\ref{eq:22}) and (\ref{eq:23});
		\STATE Initialize $ \mathbf{Q} $ and $ a $;
		\REPEAT
		\STATE Update $ \mathbf{b}$ by (\ref{eq:29});
		
			\STATE \textbf{Case 1: Amplitude Only}
		\IF {$ q\in [0,1] $}
		\STATE Update $ \mathbf{Q} $ and $ a $ by (\ref{34});
		\ENDIF
		
		\STATE \textbf{Case 2: Binary Amplitude}
		\IF {$ q = 0 $ or $ 1 $}
		\STATE Update $ \mathbf{Q} $ and $ a $ by (\ref{36}) and \textbf{Algorithm 1}, respectively;
		\ENDIF
		
		\STATE \textbf{Case 3: Lorentzian-constrained Phase}
		\IF {$ q = \frac{j+e^{j\phi}}{2}, \phi\in[0,2\pi] $}
		\STATE Update $ \mathbf{Q} $ and $ a $ by (\ref{43}) and (\ref{42}), respectively;
		\ENDIF
		
		\STATE Normalize $ \mathbf{b}$ by (\ref{44});
		\STATE Caculate $ R_{\mathrm{min}} $ by (\ref{eq:object1});
		\UNTIL $ R_{\mathrm{min}} $ converges
		\STATE Caculate $\mathbf{b} $ and $ \mathbf{Q} $ by (\ref{eq:45});
		\STATE Return $ R_{\mathrm{min}} $, $\mathbf{b} $ and $ \mathbf{Q} $. 
	\end{algorithmic}
\end{algorithm}

\end{itemize}

\subsection{Convergence and Complexity Analysis}
\subsubsection{Convergence}
We briefly explain the convergence of the proposed algorithm. As shown in \textbf{Algorithm 2}, the fully digital relaxed solution obtained is in closed form, so it is only necessary to ensure the convergence of the subsequent norm approximation iterations. Specifically, the value of $ O_{e}(\mathbf{Q},\mathbf{b}) $ is greater than or equal to zero, based on the non-negativity of the norm. Then, we introduce the superscript \(t\) as the iteration index, and the convergence of the iterations can be summarized as follows:
\begin{align}
	O_{e}(\mathbf{Q}^{(t)},\mathbf{b}^{(t)})&\overset{(a)}\geq O_{e}(\mathbf{Q}^{(t)},\mathbf{b}^{(t+1)}) \overset{(b)}\geq O_{e}(\mathbf{Q}^{(t+1)},\mathbf{b}^{(t+1)}).
\end{align}
where (a) and (b) follow since the updates of digital beamformer $ \mathbf{b} $ in (\ref{eq:29}) and analog beamformer $ \mathbf{Q} $ in (\ref{34}), (\ref{36}), \textbf{Algorithm 1} and (\ref{43}) are monotonous.
\subsubsection{Complexity}
The computational complexity of the proposed beamforming design scheme is mainly introduced by the updates of variables $ \mathbf{b} $ and $ \mathbf{Q} $, as shown in (\ref{eq:29}), (\ref{34}), (\ref{36}), \textbf{Algorithm 1}, and (\ref{43}), respectively. Let \( N_{RF} \) and \( N_t \) denote the number of RF chains and transmit antennas, respectively. The computational complexity of obtaining the fully-digital beamformer \( \mathbf{f} \) is \( \mathcal{O}((K+L)N_t)\) according to {\it Proposition 1}. Furthermore, the computational complexity of updating \( \mathbf{b} \) is \( \mathcal{O}(N_{RF}N_t + N_{RF}^3) \), primarily attributed to the matrix inversion and the unique properties of the diagonal matrix \( \mathbf{Q} \) in (\ref{eq:29}). Additionally, the three different control methods exhibit similar complexity in the analog domain. Specifically, for the amplitude only and Lorentzian-constrained phase control, the complexity of updating $ \mathbf{Q} $ in (\ref{34}) and in (\ref{43}) is both \( \mathcal{O}(N_{RF}N_t) \). Let $I_b$ denote the number of iterations in \textbf{Algorithm 1}, the complexity of updating $ \mathbf{Q} $ in (\ref{36}) and \textbf{Algorithm 1} for the Binary amplitude control is \( \mathcal{O}((N_{RF}+I_b)N_t) \). Let $ I_0 $ denote the number of iterations required by the convergence of sum-rate $ R_{\mathrm{min}} $ in \textbf{Algorithm 2}. The overall complexity of \textbf{Algorithm 2} is approximated as $ \mathcal{O}((K+L)N_t+I_0(N_{RF}N_t+N_{RF}^3)) $ for amplitude only and Lorentzian-constrained control and $ \mathcal{O}((K+L)N_t+I_0((N_{RF}+I_b)N_t+N_{RF}^3)) $ for binary amplitude control.

\section{Numerical results}
In this section, we conduct numerical results to validate the theoretical analysis of spatial orthogonality. The beam patterns of circular antenna array are also presented, and the performance of the proposed beamforming scheme in the circular H-MIMO-assisted IDET system is evaluated. As shown in Figure \ref{fig_channel}, the center of the circular antenna array is located at the origin, with the antenna plane parallel to the $ xoy $-plane. The number of transmit antennas is set to 800, and the operating frequency is set to 30 GHz \cite{10243590}. We run the simulations on a computer equipped with an AMD Ryzen 7 3700X 8-Core Processor using MATLAB R2022b.
\subsection{Verification of Spatial Orthogonality}
\begin{figure}
	\centering  
	\subfigbottomskip=1pt 
	\subfigcapskip=-2pt 
	\subfigure[]{
		\includegraphics[width=1\linewidth]{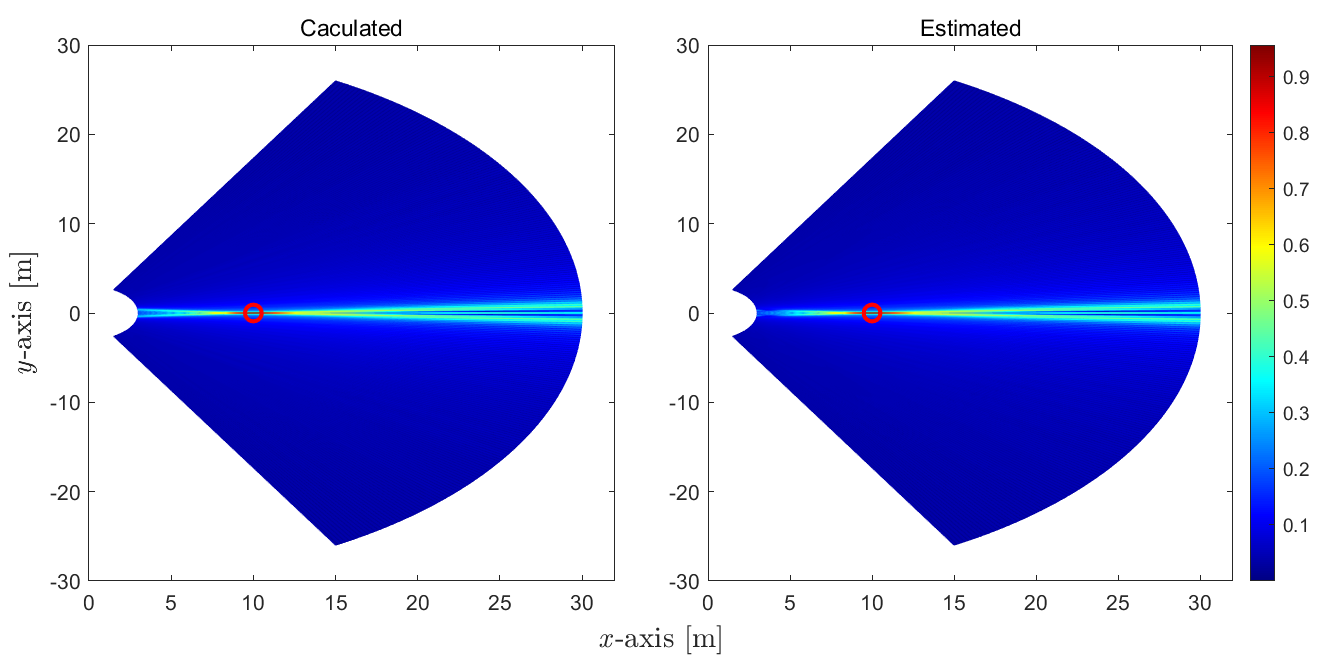} 
	}		
	\\
	\subfigure[]{
		\includegraphics[width=1\linewidth]{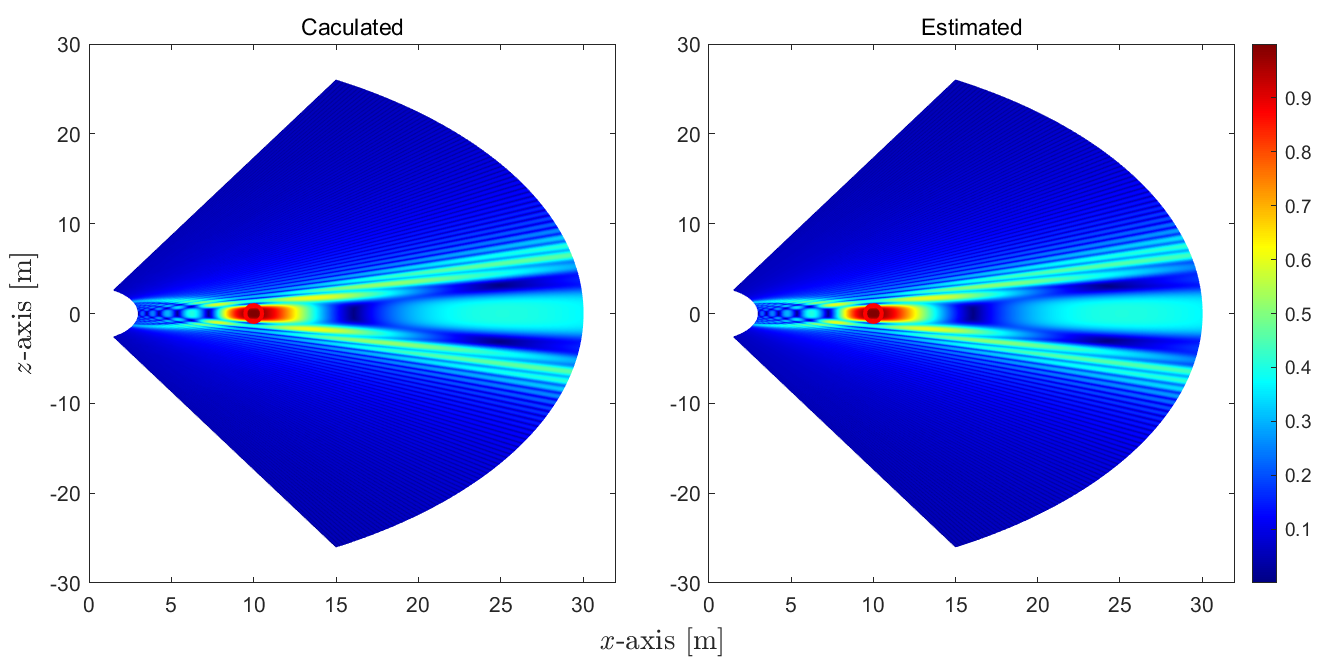} 
	}	
	\\
	\subfigure[]{
		\includegraphics[width=1\linewidth]{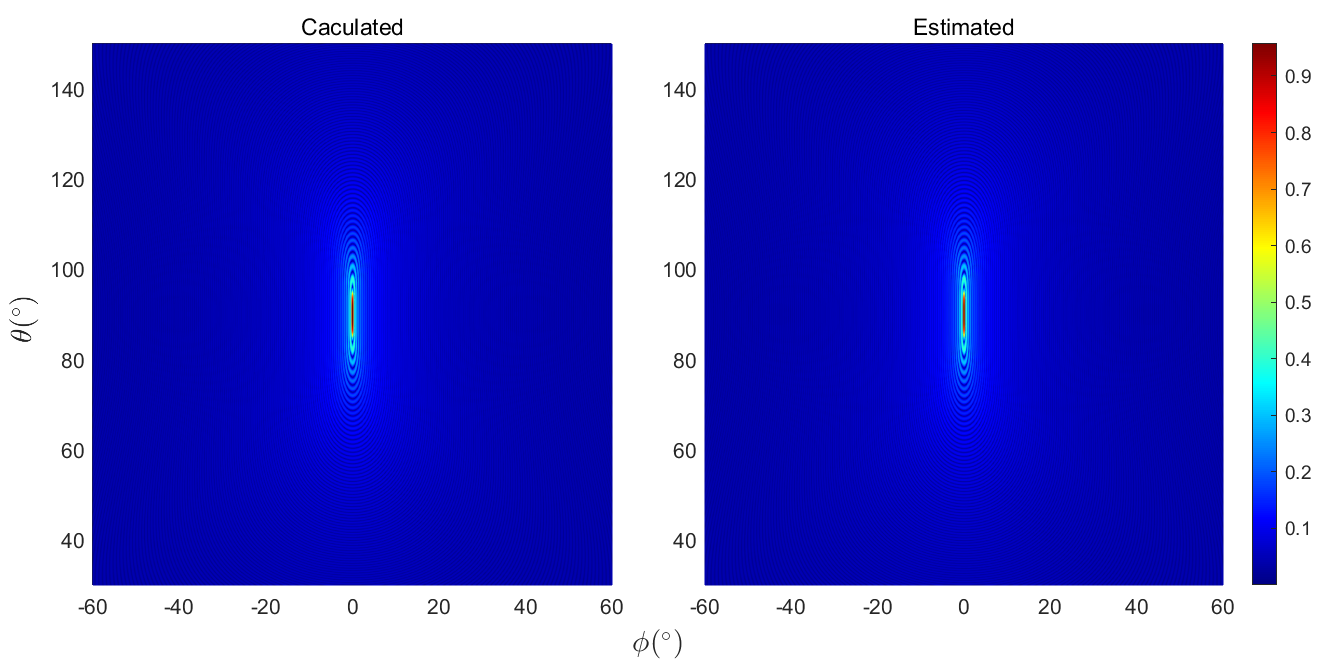} 
	}
	\caption{Validation of the resolution function in different planes: the Horizontal plane, \textit{i.e.}, $ xoy $ (a), Vertical plane, \textit{i.e.}, $ xoz $ (b), and angular domain (c). }
	\label{fig_resolution_function}
\end{figure}
Before verifying the orthogonality, we first validate the accuracy of the obtained resolution function in (\ref{eq:resolution function}). The summation in (\ref{eq:resolution function}) is carried out over $n$ in the range from -80 to 80. The user is located at (10 m, 0, 0), and the beam is focused on this position. The beam patterns on the horizontal plane through the origin ($ xoy $-plane), the vertical plane through the origin ($ xoz $-plane), and the angular domain plane are shown in Fig. \ref{fig_resolution_function} (a), (b) and (c), respectively. The beam patterns on the left are obtained through exact calculations based on definition in (\ref{eq:def_resolution}), while the beam patterns on the right are obtained from our closed-form solution in (\ref{eq:resolution function}). 

We observe from Fig.  \ref{fig_resolution_function} (a) that the estimated beam pattern in the $ xoy $-plane closely matches the exact beam pattern, with only slight discrepancies within 3 meters along the x-axis. This is because our closed-form solution is based on the classical Fresnel near-field approximation, which introduces a lower bound denoted as $ \frac{1}{2}\sqrt{\frac{D^3}{\lambda}}$. When the receiver is located within the near-field region, our closed-form solution is highly accurate. Moreover, the beam of the circular antenna array in the near field can focus on a specific position in the distance domain, unlike in the far field where energy concentration can only be achieved in the angular domain.

From Fig. \ref{fig_resolution_function} (b), we observe that the estimated beam pattern in the $ xoz $-plane is almost identical to the exact beam pattern. In addition, beam focusing can also be achieved in the $ xoz $-plane. However, the focusing effect is not as effective as in the $ xoy $-plane, as the focal point is larger and the beam is wider. 
Therefore, in practical deployments, since users are mostly located within the same plane, to achieve better beam focusing, the circular antenna array should be placed parallel to the horizontal plane.

From Fig. \ref{fig_resolution_function} (c), we observe that the estimated beam pattern in the angular domain is also accurate. Moreover, the energy concentration capability of the circular antenna array beam varies across different angular domains. The beam width in the $ \phi $-axis is narrower than that in the $ \theta $-axis. Therefore, by utilizing antenna rotation techniques, we may achieve better beam focusing performance in the future work.

\begin{figure}[t]
	\centering
	\includegraphics [width=88mm]{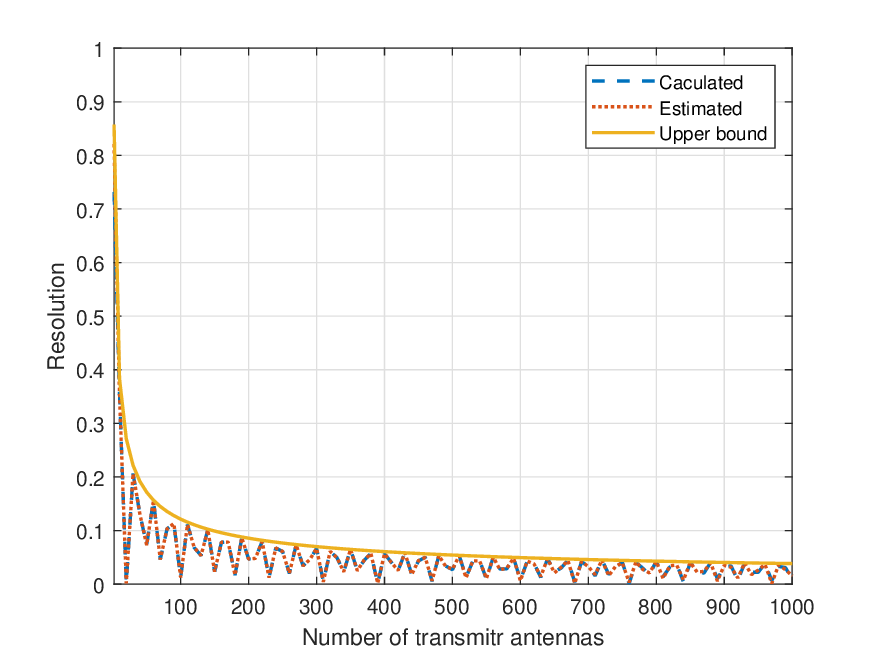}\\
	\caption{Verification of asymptotic spatial orthogonality.}
	\label{fig_orthogonality}
\end{figure}
We validate the proposed the asymptotic spatial orthogonality of channels for circular antenna array in the near-field in Fig. \ref{fig_orthogonality}. Without loss of generality, we select two points in the spherical coordinate system $(r, \theta, \phi)$: $(15 \, \text{m}, \frac{\pi}{2}, 0)$ and $(20 \, \text{m}, \frac{\pi}{6}, \frac{\pi}{3})$. In addition to the resolution function values calculated based on the definition in (\ref{eq:def_resolution}) and the resolution function values estimated using the closed-form solution in (\ref{eq:resolution function}), we also plot the upper bound value provided in (\ref{eq：resolution upper boound}) using a solid line. It can be observed from Fig. \ref{fig_orthogonality} that, for any two points in three-dimensional space, the value of the resolution function varies irregularly with the number of antennas and does not exhibit the form of a single Bessel function. However, the proposed closed-form solution remains accurate, and the obtained upper bound is tight and smooth. Moreover, as the number of antennas increases, the upper bound of the resolution function gradually approaches zero, thereby proving that the asymptotic orthogonality is correct.
\subsection{Near-field Focusing of Circular H-MIMO}
\begin{figure}[t]
	\centering
	\includegraphics [width=90mm]{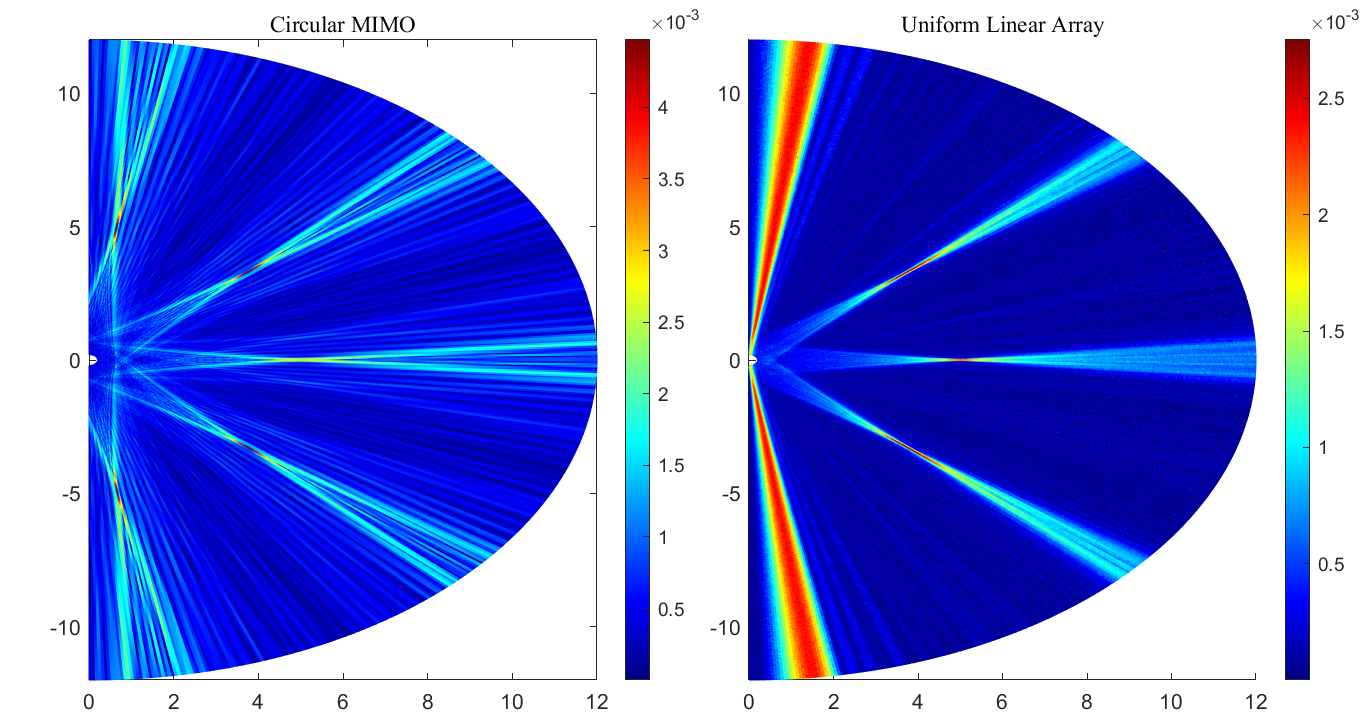}\\
	\caption{Near-field focusing comparison between circular and linear antenna arrays.}
	\label{fig_comparision_nearfield}
\end{figure}
We investigate the near-field focusing effect of the circular antenna array in Fig.(\ref{fig_comparision_nearfield}). The diameter of the circular antenna array is approximately 1.28 m, which is equivalent to the width of a uniform linear array with 256 antennas. Therefore, we use the uniform linear array with 256 antennas as the control group. Furthermore, to illustrate the impact of the antenna shape, we temporarily disregard the specific hardware constraints of H-MIMO. Assuming five DUs are located at a distance of 5 m from the center of the array at five different angles \((0, \pm \frac{11}{24}, \pm \frac{11}{48})\), the beamforming scheme can be obtained from {\it Proposition 1} by setting the number of energy user to zero. The received power at each location is normalized by the channel gain. We observe from Fig. \ref{fig_comparision_nearfield} that both the circular antenna array and the uniform linear array can achieve multi-point beam focusing. However, when the user's position deviates significantly from the central axis of the linear array, with the absolute value of \(\phi\) approaching \(90^\circ\), the beam focusing effect of the linear array degrades into far-field beam steering. In contrast, the circular antenna array can maintain consistent focusing performance at different angles. This is because the circular antenna array has the same equivalent array width at different angles in the \(\phi\)-domain, ensuring that near-field focusing is unaffected by the specific value of \(\phi\).

\begin{figure*}
	\centering  
	\subfigbottomskip=0pt 
	\subfigcapskip=-2pt 
	\subfigure[]{
		\includegraphics[width=0.47\linewidth]{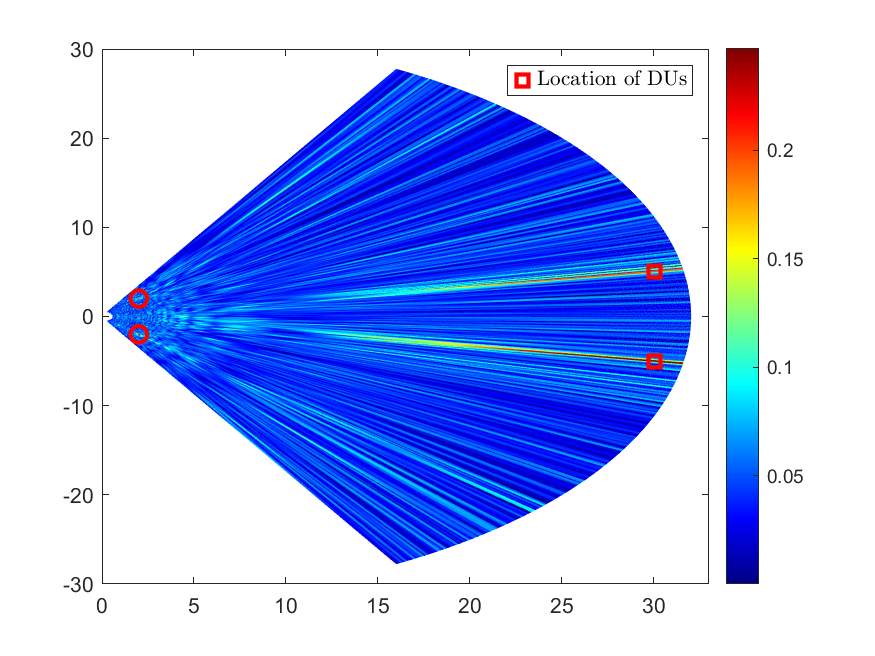} 
	}		
	\subfigure[]{
		\includegraphics[width=0.47\linewidth]{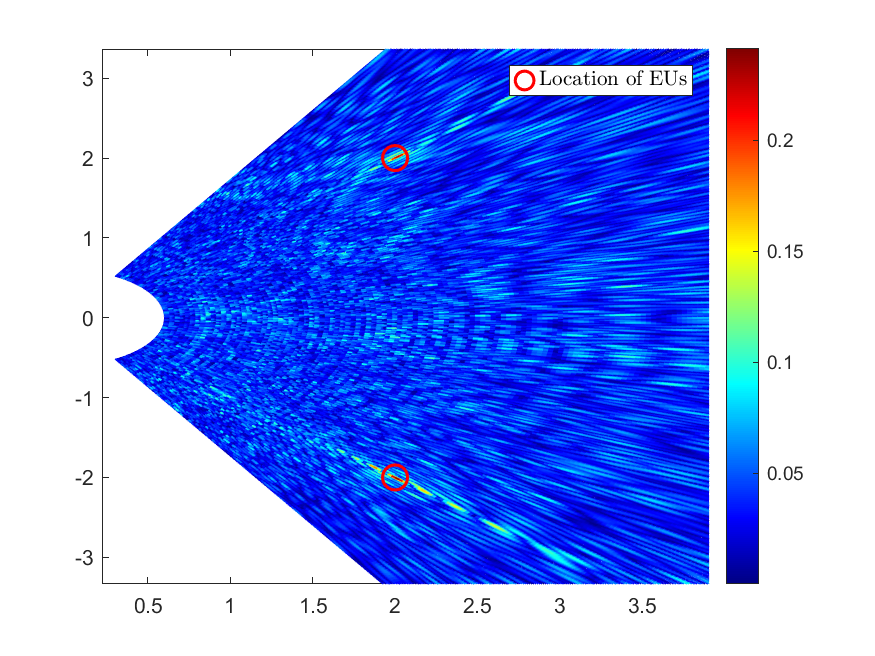} 
	}	
	\caption{Near-filed focusing for the IDET system: (a)Near-field Focusing for DUs, (b)Near-field Focusing for EUs}
	\label{fig_nearfied_focusing}
\end{figure*}
Next, we analyze the near-field focus effect for the circular H-MIMO assisted IDET system in Fig. \ref{fig_nearfied_focusing}. DUs and EUs are located in the same plane parallel to the circular antenna array but with a certain height difference. Specifically, their positions in the Cartesian coordinate system are \((30, \pm 5, -1)\) for DUs and \((2, \pm 2, -1)\) for EUs, respectively. The received power at each location is normalized by the channel gain. We can observe from Fig. \ref{fig_nearfied_focusing} that the received signal power at the user positions is marked in red, indicating that the beamforming gain is strongest at these locations. Multi-point beam focusing is successfully achieved in the circular H-MIMO assisted IDET system. Due to the hardware constraints of circular H-MIMO and the potential interference from the beams of nearby EUs on distant locations, the beam focusing in Figure \ref{fig_nearfied_focusing} exhibits more clutter compared to Fig \ref{fig_comparision_nearfield}. However, it should be noted that such a large-scale fully digital array, as used in Figure 5, does not exist in practice. Only H-MIMO can achieve such a large number of antennas.

\subsection{Performance Evaluation of Circular H-MIMO assisted IDET System}
In this subsection, we evaluate the performance of beamforming design proposed in \textbf{Algorithm 2} for the circular H-MIMO assisted IDET system. Every user is equipped with a single antenna, and the angles of all users relative to the transmitter are randomly generated. The achievable rate is defined as the minimum rate among all data users, and the other simulation parameters are listed in TABLE \ref{tab:simulation_parameters}.


\begin{table}[htbp]
	\centering
	\caption{Simulation Parameters}
	{	\small
	\begin{tabular}{|l|l|}
		\hline
		\textbf{Parameter} & \textbf{Value} \\ \hline
		Number of Transmit Antennas & 800 \cite{10243590}\\ \hline
		Operating Frequency & 30 GHz \cite{10243590}\\ \hline
		Number of Data Users (DUs) & 3 \\ \hline
		Number of Energy Users (EUs) & 2 \\ \hline
		Distance to EUs & 3 m \\ \hline
		Distance to DUs & 20 m \\ \hline
		Transmit Power \( P_t \) & 20 dBm \\ \hline
		Energy Harvesting Requirement \( E_0 \) & -15 dBm \cite{9454368} \\ \hline
		Number of Paths \( I \) in Eq. \ref{eq:channel} & 6 \\ \hline
		Noise Power \( \sigma^2 \) & -94 dBm \\ \hline
		Transmit Antenna Gain & 10 dBi \\ \hline
		Waveguide Attenuation Coefficient \( \gamma \) & 5 \\ \hline
	\end{tabular}}
	\label{tab:simulation_parameters}
\end{table}

In Eq. (31), Eq. (32), and Eq. (33), the original beamforming schemes, designed for the three control methods: Amplitude only, Binary amplitude, and Lorentzian-constrained phase, are denoted as `Amplitude,' `0-1,' and `Phase,' respectively. As shown in \textbf{Algorithm 2}, the equivalently scaled and improved beamforming schemes are represented as `Amplitude scaling,' `0-1 scaling,' and `Phase scaling,' respectively. For the performance comparison, we consider the following benchmark schemes:
\begin{itemize}	
	\item {\bf Upper bound:} In Proposition 1, the optimal, fully digital beamforming scheme derived using asymptotic orthogonality is presented, denoted as ``FD ASY.'' This can be considered the unattainable upper bound of our proposed scheme, as the practical hardware constraints of H-MIMO prevent it from achieving full control in the complete complex domain.
	\item {\bf Genetic algorithm (GA):} In the design of the equivalently scaled beamforming scheme for Lorentzian-constrained phase-controlled circular H-MIMO, we obtain a low-complexity suboptimal solution to the single-variable non-convex optimization problem in Eq.~(40). To validate the effectiveness of this suboptimal solution, we use a GA to solve Eq.~(40) as a comparison, denoted as ``Phase SGA.''
	\item {\bf Match-filtering (MF):} Yuan et al. \cite{9139999} proposed an MF scheme for the uplink of an ideal continuous surface H-MIMO system without considering hardware constraints, aiming to maximize the target signal of users. We extend this scheme to the downlink scenario, treating DUs and EUs equally. Since hardware constraints are not considered, this scheme can also be regarded as a fully digital beamforming scheme, which we denote as ``FD MF''. While it has the same low complexity as our proposed scheme, it does not account for the trade-off between communication and energy transfer or the fair resource allocation among users.
	\item {\bf Successive convex approximation (SCA):} For problem \( P_2 \), a common solution approach is to use SCA combined with convex approximation methods such as first-order Taylor expansion. This approach generally achieves good performance but comes with relatively high complexity and requires iterative optimizing. Although this method does not account for hardware constraints,  it can be regarded as the performance upper bound of traditional SCA-based schemes for comparison, denoted as ``FD SCA.''
\end{itemize}

\begin{figure}[t]
	\centering
	\includegraphics [width=88mm]{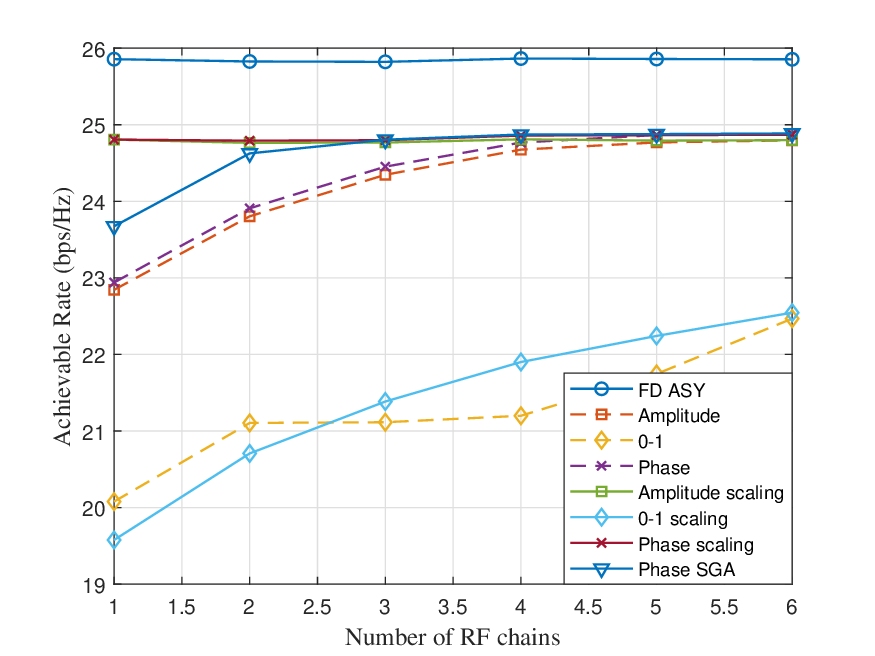}\\
	\caption{Achievable rate of DUs versus number of RF chains.}
	\label{fig_RF_chain}
\end{figure}
We characterize the impact of the number of RF chains on the achievable rate in Fig. \ref{fig_RF_chain}. We can see from Fig. \ref{fig_RF_chain} that the fully digital asymptotically optimal beamforming scheme ``FD ASY'' requires only one RF chain to achieve the best performance. This is because, in a multicast system, we transmit only one identical signal, and the fully digital transmitter has sufficient control freedom to implement the optimal beamforming. 

For H-MIMO systems with Amplitude only and Lorentzian-constrained phase control mode, the achievable rate of the original beamforming scheme ``Phase'' and ``Amplitude'' gradually increases with the number of RF chains. This is because, due to practical hardware limitations, H-MIMO lacks sufficient control freedom, while more RF chains provide greater control freedom to compensate for the hardware constraints in the analog domain. When the number of RF chains is sufficiently large, the achievable rate of the original H-MIMO converges to a nearly same constant value. Additionally, the equivalently scaled and improved beamforming scheme ``Phase scaling'' and ``Amplitude scaling'' reaches the converged value with just one RF chain. This is because the proposed improvement relaxes the hardware constraints in the analog domain, and after obtaining the converged solution, it is transformed back into a feasible solution that satisfies the constraints, thereby increasing the control freedom. Additionally, the improved scheme ``Phase SGA'' solved by the genetic algorithm, although superior to the original scheme ``Phase'', still performs worse than the proposed ``Phase scaling''. This demonstrates the effectiveness of our low-complexity solution.

For the Binary amplitude control scheme of H-MIMO, both the original scheme ``0-1'' and the improved scheme ``0-1 scaling'' show an increase in the achievable rate as the number of RF chains increases. When the number of RF chains is small, ``0-1'' outperforms ``0-1 scaling''. However, as the number of RF chains increases, ``0-1 scaling'' gradually surpasses ``0-1''. This is because the binary amplitude control scheme has very limited control freedom, and the norm-based approximation scheme does not provide a good solution. As the number of RF chains increases, the control freedom in the digital domain increases, and the norm approximation leads to better solutions, thereby improving the performance of ``0-1 scaling'' significantly.

\begin{figure}[t]
	\centering
	\includegraphics [width=88mm]{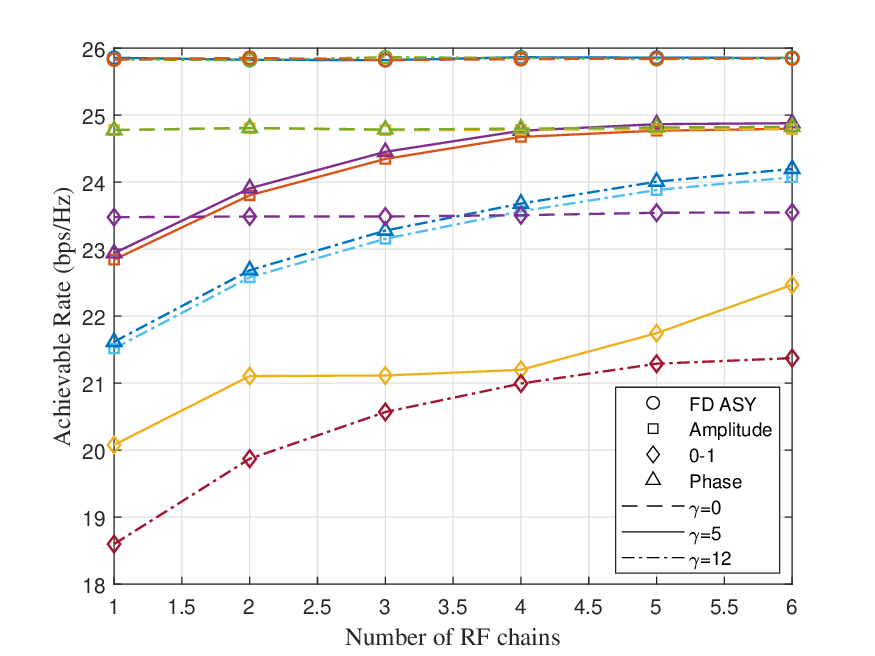}\\
	\caption{Achievable rate of DUs versus number of RF chains.}
	\label{fig_RF_chain_loss_factor}
\end{figure}	
We investigate the impact of the waveguide attenuation coefficient \(\gamma\) in Fig. \ref{fig_RF_chain_loss_factor}. We observe from Fig. \ref{fig_RF_chain_loss_factor} that, when the attenuation coefficient \(\gamma = 5\), the performance of all schemes is consistent with that in Fig. 7. When \(\gamma = 0\), the achievable rate of the original scheme with just one RF chain also converges to the same rate as the improved scheme. The smaller the attenuation coefficient \(\gamma\), the fewer RF chains are required for convergence. This indicates that when the waveguide attenuation coefficient can be neglected, the original scheme can also achieve good performance.

\begin{figure}[t]
	\centering
	\includegraphics [width=88mm]{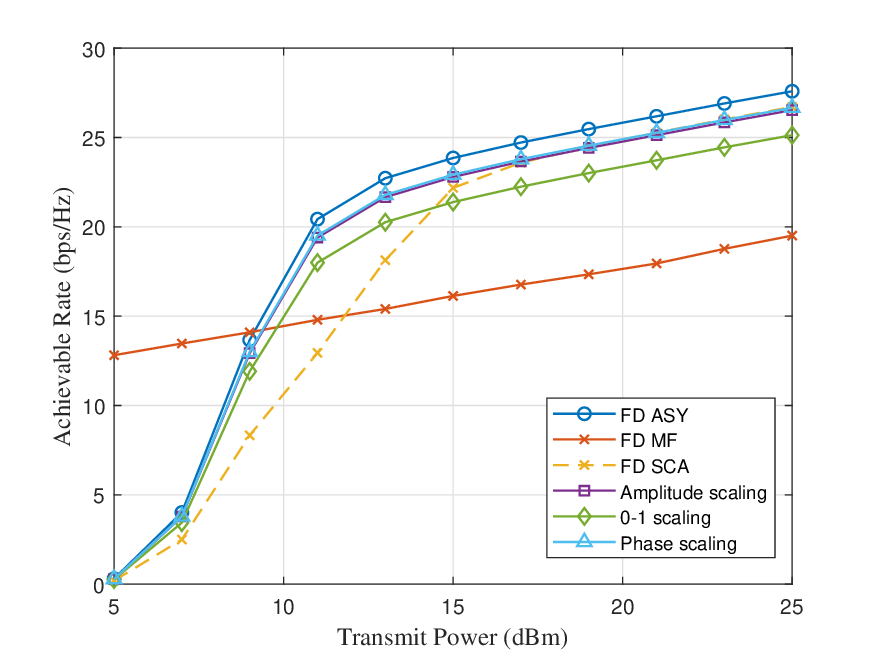}\\
	\caption{Achievable rate of DUs versus transmit power.}
	\label{fig_txpower}
\end{figure}
We characterize the impact of transmit power on the achievable rate in Fig. \ref{fig_txpower}, which shows that the achievable rate increases continuously with increasing transmit power. The spatial orthogonality based fully digital scheme ``FD ASY'' performs the best, but the full digital architecture is not practical for large-scale MIMO, which can be considered as the performance upper bound. Both ``Amplitude scaling'' and ``Phase scaling'' achieve almost the same achievable rate, while the rate of ``0-1 scaling'' is slightly lower. This is because both Amplitude-only and Lorentzian-constrained phase schemes have similar control spaces in the half-complex plane, while binary amplitude has much fewer degrees of freedom for control. Nevertheless, binary amplitude remains the easiest to implement and enables very fast control. In addition, as the transmit power increases, both the ``amplitude scaling'' and ``phase scaling'' maintain a consistent gap from the upper bound, which is approximately 0.85 bps/Hz. Moreover, the achievable rate of ``FD SCA'' is lower than that of ``Phase scaling'' at low transmit power, but at high power, it becomes almost the same as ``Phase scaling''. Despite the use of a fully digital architecture, the high-complexity traditional scheme only achieves the same performance as our proposed solutions, which demonstrates the effectiveness of the proposed schemes. Furthermore, ``FD MF'' has a similar complexity to the proposed scheme but lacks a carefully designed allocation of spatial and power resources, leading to the poorest performance.

\begin{figure}[t]
	\centering
	\includegraphics [width=88mm]{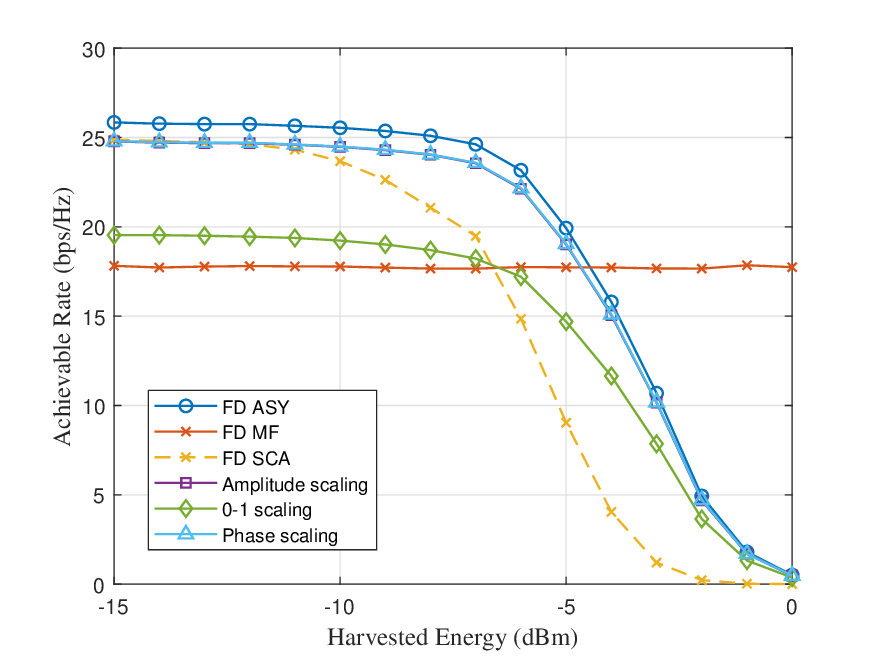}\\
	\caption{Harvested Energy of EUs versus achievable rate of DUs.}
	\label{fig_energy}
\end{figure}
\begin{figure}[t]
	\centering
	\includegraphics [width=88mm]{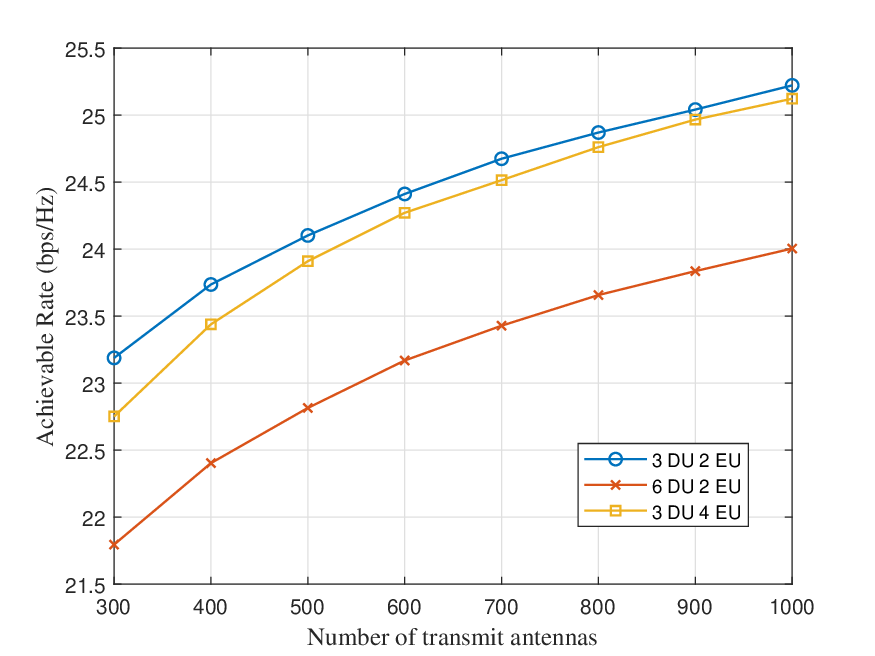}\\
	\caption{Achievable rate of DUs versus number of transmit antennas.}
	\label{fig_txantenna}
\end{figure}
We investigate the impact of energy harvesting demand of energy users on the achievable rate in Fig. \ref{fig_energy}. We notice from Fig. \ref{fig_energy} that the achievable rate decreases with increasing energy harvesting demand. This is because more transmit power is allocated to EUs, reducing the achievable rate for DUs. This demonstrates that the trade-off between WDT and WET is effectively realized by the proposed scheme. Additionally, consistent with the results in Fig. \ref{fig_txpower}, ``FD ASY'' performs the best, followed by ``Amplitude scaling'' and ``Phase scaling'', which outperform ``0-1 scaling''. The high-complexity ``FD SCA'' achieves the same performance as the proposed schemes at low energy harvesting demand, but lags behind at higher energy demand. Furthermore, the achievable rate of ``FD MF'' remains almost constant across different energy harvesting demands, as it does not consider the joint design of WDT and WET.

\begin{table}[htbp]
	\small
	\centering
	\caption{Average Runtime of Different Algorithms}
	\label{tab:runtime}
	\begin{tabular}{|l|c|}
		\hline
		\textbf{Algorithm} & \textbf{Average Runtime (s)} \\
		\hline
		FD ASY             & 0.000132      \\
		FD MF              & 0.000073      \\
		Amplitude Scaling  & 0.014511      \\
		0-1 Scaling        & 0.038685      \\
		Phase Scaling      & 0.018220      \\
		FD SCA             & 4.041102      \\
		\hline
	\end{tabular}
\end{table}
We characterize the average running times of different schemes in TABLE \ref{tab:runtime}. We observe that the Amplitude Scaling, 0-1 Scaling and Phase Scaling schemes exhibit significantly lower computational complexity compared to FD SCA, as their iterations are based on closed-form solutions. Among them, 0-1 Scaling involves iterative updates from {\bf Algorithm 2}, resulting in relatively higher complexity. Moreover, although FD MF has the shortest runtime, it fails to achieve a trade-off between WDT and WET, and delivers the worst performance.

We investigate the impact of the number of transmitting antennas on the achievable rate in Fig. \ref{fig_txantenna}. We observe that the achievable rate increases with the number of transmitting antennas. This is because more antennas provide a greater spatial gain. Furthermore, as the number of EUs or DUs increases, the achievable rate decreases. This is because additional users consume the power and spatial resources originally allocated to other users, resulting in a reduction in the minimum achievable rate.
\section{Conclusions and  Future Directions}
In this paper, we have developed beamforming designs for the circular H-MIMO-assisted integrated data and energy transfer system under various control modes. First, by deriving the closed-form near-field resolution function, we have shown the asymptotic spatial orthogonality of the near-field channel. Based on the orthogonality, we then have designed beamforming schemes that maximize the minimum achievable rate of data users while meeting energy harvesting requirements of energy users. The proposed schemes, with low complexity, outperform benchmark schemes and effectively balance wireless data transfer and wireless energy transfer. Numerical results have confirmed the validity of the resolution function and asymptotic spatial orthogonality, with circular H-MIMO achieving a broader angular range than traditional linear MIMO systems. Furthermore, the improved scheme, achieved through scaling of the analog domain, requires only a single RF chain to achieve the saturated achievable rate that the original scheme attains when equipped with a sufficient number of RF chains.

Due to the presence of a large number of antenna elements, acquiring perfect channel state information in H-MIMO systems is highly challenging. In the future, exploring robust beamforming design becomes a promising research direction, where techniques such as meta learning and manifold learning may be employed \cite{10623434}. In addition, it is an interesting direction to study beamforming in mobile scenarios. Related research topics include beam tracking mechanisms based on user motion or angle prediction \cite{10319287}, temporal CSI prediction methods \cite{10417075}, and wideband waveform designs such as orthogonal time frequency space (OTFS) that compensate for or exploit Doppler shifts \cite{10632049}.

\begin{appendices}
\section{Proof of Lemma 1}
Considering the distance from the transmitter to a user is much larger than the aperture of the transmitter, \textit{i.e.}, $ r \gg R $, we ignore the  last two terms of (\ref{eq:3}). By utilizing the approximation in (\ref{eq:3}), we obtain (\ref{eq:resolution derivation}),  as shown at the top of the next page,
\begin{figure*}[ht]
	\centering
	\begin{align}\label{eq:resolution derivation}
		&\Delta(r_1,\theta_1,\phi_1,r_2,\theta_2,\phi_2) \nonumber\\  \approx& \left|\frac{1}{2\pi}\sum_{n=1}^{N}e^{j\frac{2\pi}{\lambda}R\left(\sin\theta_2\cos(\phi_2-\psi_n) -\sin\theta_1\cos(\phi_1-\psi_n)+\frac{R}{2r_1}\left( 1-\cos^2(\phi_1-\psi_n)\sin^2\theta_1\right)- \frac{R}{2r_2}\left( 1-\cos^2(\phi_2-\psi_n)\sin^2\theta_2\right)\right) }  \right|\nonumber\\
		\approx& \left|\frac{1}{2\pi}\int_{0}^{2\pi}e^{j\frac{2\pi}{\lambda}R\left(\sin\theta_2\cos(\phi_2-x) -\sin\theta_1\cos(\phi_1-x)+\frac{R}{2r_1}\left( 1-\cos^2(\phi_1-x)\sin^2\theta_1\right)- \frac{R}{2r_2}\left( 1-\cos^2(\phi_2-x)\sin^2\theta_2\right)\right) } \mathrm{d}x \right| \nonumber\\
		=&\Bigg|\frac{1}{2\pi}\int_{0}^{2\pi}e^{j\frac{2\pi}{\lambda}R \left(\left( \sin\theta_2\cos\phi_2-\sin\theta_1\cos\phi_1\right) \cos x -\left( \sin\theta_2\sin\phi_2-\sin\theta_1\sin\phi_1\right) \sin x \right)} \nonumber \\  
		&\times e^{j\frac{2\pi}{\lambda}R^2 \left(\left( \frac{\sin^2\theta_2}{4r_2}\cos(2\phi_2)-\frac{\sin^2\theta_1}{4r_1}\cos(2\phi_1)\right)\cos(2x)+\left( \frac{\sin^2\theta_2}{4r_2}\sin(2\phi_2)-\frac{\sin^2\theta_1}{4r_1}\sin(2\phi_1)\right)\sin(2x)+\frac{\sin^2\theta_2}{4r_2}-\frac{\sin^2\theta_1}{4r_1}+\frac{1}{2r_1}-\frac{1}{2r_2} \right) }\mathrm{d}x \Bigg|\nonumber \\  
		=&\Bigg|\frac{\xi_5}{2\pi}\int_{0}^{2\pi}e^{j\left( \sqrt{\eta_1^2+\eta_2^2}\sin(x+\arctan\frac{\eta_1}{\eta_2} )+\sqrt{\eta_3^2+\eta_4^2}\sin(2x+\arctan\frac{\eta_3}{\eta_4} )\right) }\mathrm{d}x \Bigg|
	\end{align}
	\vspace*{8pt}
	\hrulefill
\end{figure*}

Then, we have
\begin{align}
	&\Delta(r_1,\theta_1,\phi_1,r_2,\theta_2,\phi_2) 
	=\Bigg|\frac{\xi_5}{2\pi}\int_{0}^{2\pi}e^{j\left( \xi_1\sin(x+\xi_2 )+\xi_3\sin(2x+\xi_4 )\right) }\mathrm{d}x \Bigg| \nonumber \\
	\overset{\text{(a)}}=&\left| \xi_5 I_0(j\xi_1)I_0(j\xi_3)+2\xi_5\sum_{n=1}^{\infty}I_n(j\xi_3)I_{2n}(j\xi_1)e^{jn(2\xi_2-\xi_4-\pi/2)} \right| \nonumber \\
	=& \left| \xi_5 J_0(\xi_1)J_0(\xi_3)+2\xi_5\sum_{n=1}^{\infty}j^{-n}J_n(-\xi_3)j^{-2n}J_{2n}(-\xi_1)e^{jn(2\xi_2-\xi_4-\pi/2)}\right| \nonumber \\
	=&\left| \xi_5\sum_{n=-\infty}^{\infty}j^n J_n(\xi_3)J_{2n}(\xi_1)e^{jn(2\xi_2-\xi_4-\pi/2)}\right| 
\end{align}
where (a) is realized by utilizing $ \frac{1}{2\pi}\int_{0}^{2\pi}e^{ z\cos(m\theta)+y\cos\theta }\mathrm{d}\theta=I_0(z)I_0(y)+2\sum_{n=1}^{\infty}I_n(z)I_{mn}(y) $.

\section{Proof of Lemma 2}
\begin{align}\label{eq：resolution upper boound}
	&\mathbf{a}_t^H(r_1,\theta_1,\phi_1)\mathbf{a}_t(r_2,\theta_2,\phi_2)\nonumber \\
	&\leq \Delta(r_1,\theta_1,\phi_1,r_2,\theta_2,\phi_2) \nonumber \\
	& =\left| \xi_5\sum_{n=-\infty}^{\infty}j^n J_n(\xi_3)J_{2n}(\xi_1)e^{jn(2\xi_2-\xi_4-\pi/2)}\right| \nonumber \\
	&\overset{\text{(a)}}\approx \left| \xi_5\sqrt{\frac{2}{\pi\xi_1}}\sum_{n=-\infty}^{\infty}j^n J_n(\xi_3)\cos(\xi_1-n\pi-\pi/4)e^{jn(2\xi_2-\xi_4-\pi/2)}\right| \nonumber \\
	&= \left| \xi_5\cos(\xi_1-\pi/4)\sqrt{\frac{2}{\pi\xi_1}}\sum_{n=-\infty}^{\infty}(-je^{j(2\xi_2-\xi_4-\pi/2)})^n J_n(\xi_3)\right| \nonumber \\
	&\overset{\text{(b)}}=\left| \xi_5\cos(\xi_1-\pi/4)\sqrt{\frac{2}{\pi\xi_1}} e^{-j\xi_3\sin(2\xi_2-\xi_4)}\right|
	\nonumber \\
	&\leq\sqrt{\frac{2}{\pi\xi_1}}
\end{align}
where (a) is realized by utilizing the asymptotic property of $ J_n(\cdot) $ in Eq.(9.2.1) of \cite{mabramowitz64:handbook}. (b) is realized by utilizing Eq.(9.1.41) in \cite{mabramowitz64:handbook}. Supposing that $ \theta_1 \neq \theta_2 $ or $ \phi_1 \neq \phi_2 $, we have

\begin{align}
	&\lim_{\substack{N \rightarrow \infty}} \mathbf{a}_t^H(r_1,\theta_1,\phi_1)\mathbf{a}_t(r_2,\theta_2,\phi_2)\leq\lim_{\substack{N \rightarrow \infty}}\sqrt{\frac{2}{\pi\xi_1}}\nonumber 
\end{align}
\begin{align}
	&\approx \lim_{\substack{N \rightarrow \infty}}\sqrt{\frac{4}{\pi N\sqrt{ \sin^2\theta_1+\sin^2\theta_2-2\sin\theta_1\sin\theta_2\cos(\phi_1-\phi_2) } }}
	\nonumber \\
	&=0.
\end{align}
When $ \theta_1 =\theta_2 $, $ \phi_1 = \phi_2 $ and $ r_1 \neq r_2 $, we have
\begin{align}
	&\lim_{\substack{N \rightarrow \infty}}\mathbf{a}_t^H(r_1,\theta,\phi)\mathbf{a}_t(r_2,\theta,\phi)\nonumber \\
	&\leq \lim_{\substack{N \rightarrow \infty}} \Delta(r_1,r_2,\theta,\phi) \nonumber \\
	& =\lim_{\substack{N \rightarrow \infty}}\left| \xi_5 \int_{0}^{2\pi}e^{j\left(\frac{\pi R^2 \sin^2\theta}{2\lambda}(\frac{1}{r_2}-\frac{1}{r_1})\cos(2x-2\phi )\right) }\mathrm{d}x \right| \nonumber \\
	& \approx\lim_{\substack{N \rightarrow \infty}}\left| \xi_5 J_0(\frac{ \lambda N^2 \sin^2\theta(r_1-r_2)}{32 \pi r_1 r_2}) \right|  =0
\end{align}
Therefore, when $ \theta_1 \neq\theta_2 $, $ \phi_1 \neq \phi_2 $ or $ r_1 \neq r_2 $, we have $ \lim_{\substack{N \rightarrow \infty}} \mathbf{a}_t^H(r_1,\theta_1,\phi_1)\mathbf{a}_t(r_2,\theta_2,\phi_2)=0 $.

\section{Proof of Proposition 1}
\begin{align}
	&\lim_{\substack{N \rightarrow \infty}}\frac{1}{N}\mathbf{h_1}^H\mathbf{h_2}\nonumber \\=&	\lim_{\substack{N \rightarrow \infty}}\frac{1}{N}\left( 	\sqrt{N}\alpha_1^{[0]} \mathbf{a}_t(r_1^{[0]},\theta_1^{[0]},\phi_1^{[0]})+\sqrt{\frac{N}{I}}\sum_{i=1}^{I}\alpha^{[i]} \mathbf{a}_t(r_1^{[i]},\theta_1^{[i]},\phi_1^{[i]})\right)^H \nonumber \\ &\times \left( 	\sqrt{N}\alpha_2^{[0]} \mathbf{a}_t(r_2^{[0]},\theta_2^{[0]},\phi_2^{[0]})+\sqrt{\frac{N}{I}}\sum_{i=1}^{I}\alpha_2^{[i]} \mathbf{a}_t(r_2^{[i]},\theta_2^{[i]},\phi_2^{[i]})\right)\nonumber \\ =&\lim_{\substack{N \rightarrow \infty}}(\alpha_1^{[0]})^H\alpha_2^{[0]}\mathbf{a}_t^H(r_1^{[0]},\theta_1^{[0]},\phi_1^{[0]}) \mathbf{a}_t(r_2^{[0]},\theta_2^{[0]},\phi_2^{[0]})\nonumber \\&+\sqrt{\frac{1}{I}}\sum_{i=1}^{I}(\alpha_1^{[0]})^H\alpha_2^{[i]}\mathbf{a}_t^H(r_1^{[0]},\theta_1^{[0]},\phi_1^{[0]}) \mathbf{a}_t(r_2^{[i]},\theta_2^{[i]},\phi_2^{[i]})\nonumber \\&+\sqrt{\frac{1}{I}}\sum_{i=1}^{I}(\alpha_1^{[i]})^H\alpha_2^{[0]}\mathbf{a}_t^H(r_1^{[i]},\theta_1^{[i]},\phi_1^{[i]}) \mathbf{a}_t(r_2^{[0]},\theta_2^{[0]},\phi_2^{[0]})\nonumber \\&+\frac{1}{I}\sum_{i=1}^{I}\sum_{j=1}^{I}(\alpha_1^{[i]})^H\alpha_2^{[j]}\mathbf{a}_t^H(r_1^{[i]},\theta_1^{[i]},\phi_1^{[i]}) \mathbf{a}_t(r_2^{[j]},\theta_2^{[j]},\phi_2^{[j]})\nonumber \\&=0. 
\end{align}
\section{Proof of Lemma 2}
According to {\it Theorem} 1, $ \{\mathbf{h}_k| k \in \mathcal{D}\} $ and $ \{\mathbf{h}_l| l \in \mathcal{E}\} $ are asymptotically orthogonal to one another. Then the optimal solution to (P9) can be expressed as 
\begin{equation}
	\mathbf{f}=\sum_{k=1}^{K}\omega_k \mathbf{h}_k + \sum_{l=K+1}^{K+L}\omega_l \mathbf{h}_l+\sum_{i}\tau_i\mathbf{g}_i ,
\end{equation}
where $ \mathbf{g}_i $ are vectors orthogonal to {both} $ \{\mathbf{h}_k| k \in \mathcal{D}\} $ and $ \{\mathbf{h}_l| l \in \mathcal{E}\} $.
Then, the asymptotically received signal of the $ k $-th DU and the energy harvesting power of the $ l $-th EU can be expressed as 
\begin{align}\label{eq:25}
	\lim_{ N  \rightarrow \infty }y_k&=\lim_{ N  \rightarrow \infty }\sqrt{P_t}\mathbf{h}^H_k\mathbf{f}s+n_k\nonumber \\ 
	& =\lim_{ N  \rightarrow \infty }\sqrt{P_t}\mathbf{h}^H_k\left( {\sum_{j=1}^{K}\omega_j \mathbf{h}_j} + \sum_{l=K+1}^{K+L}\omega_l \mathbf{h}_l+\sum_{i}\tau_i\mathbf{g}_i\right) s+n_k \nonumber \\ 
	& \overset{\text{(a)}}=\underbrace{\sqrt{P_t}\omega_k \mathbf{h}^H_k \mathbf{h}_k s}_{\varrho}+o(\varrho)+n_k, 
\end{align}
\begin{align}\label{eq:26}
	\lim_{ N  \rightarrow \infty }E_l&=\lim_{ N  \rightarrow \infty }||y_l||^2=\lim_{ N  \rightarrow \infty }P_t||\mathbf{h}^H_l\mathbf{f}||^2 \nonumber \\
	& = \lim_{ N  \rightarrow \infty }P_t\left\| \mathbf{h}^H_l\left( \sum_{k=1}^{K}\omega_k \mathbf{h}_k + {\sum_{j=K+1}^{K+L}\omega_j} \mathbf{h}_j+\sum_{i}\tau_i\mathbf{g}_i\right)\right\| ^2 \nonumber \\ 
	&\overset{\text{(a)}} =\underbrace{P_t\left\| \omega_l \mathbf{h}^H_l \mathbf{h}_l \right\| ^2}_{\varrho}+o(\varrho),
\end{align}
where $ f(y)\sim o(y) $ indicates that $ \lim_{y \rightarrow \infty }(f(y)/y)=0 $ and $\text{(a)} $ is realized by using $\mathbf{a}_t^H(r_1,\theta_1,\phi_1)\mathbf{a}_t(r_1,\theta_1,\phi_1)=1 $. Since the third term of $ \mathbf{f} $ has no contribution to both DUs and EUs from Eq. (\ref{eq:25}) and Eq. (\ref{eq:26}), we obtain $ \tau_i=0 $. 

Then, the asymptotically optimal digital beamformer has a form of
\begin{equation}\label{eq:21}
	\mathbf{f} = \sum_{k=1}^{K}\omega_k \mathbf{h}_k + \sum_{l=K+1}^{K+L}\omega_l \mathbf{h}_l, k \in \mathcal{D}, l \in \mathcal{E}.
\end{equation} 
The asymptotically SNR of the $ k $-th DU can be expressed as
\begin{align}
	\lim_{N \rightarrow \infty  }\gamma_k&=\lim_{N \rightarrow \infty }\frac{P_t||\omega_k \mathbf{h}^H_k \mathbf{h}_k||^2 }{\sigma^2}
\end{align}

Without loss of generality, let the $ i $-th DU have the lowest SNR, while letting the $ j $-th DU has the highest SNR. Then, we have $ \gamma_i < \gamma_j, \forall j \neq i, j \in \{1,2,\cdots,K \} $. In order to obtain a higher fair SNR of the system, we may reduce the highest SNR of the $ j $-th DU and allocate more power to the $ i $-th DU. In this way, a new solution achieving a higher fair SNR is obtained by the following process expressed as  
\begin{align}
	\omega_i'=\omega_j'=\sqrt{\frac{\omega_i^2+\omega_j^2}{2}}	\ \ \mathrm{if} \, \omega_i\neq \omega_j         
\end{align}

If there still exists a DU whose SNR is much lower than others in this new solution, the above process can be repeated to obtain a better solution. Finally, the solution converges to a situation that all the users have nearly the same SNR. In other words, the asymptotically optimal $ 	\mathbf{f} $ is obtained when all DUs have the same achievable rate \cite{10559446}. By considering the constraint of (\ref{eq:object2}a), we have the following equation
\begin{equation}
	\begin{cases}
		\left\| \sum_{k=1}^{K}\omega_k \mathbf{h}_k + \sum_{l=K+1}^{K+L}\omega_l \mathbf{h}_l\right\|^2 = 1, \\
		R_i=R_j, \forall i,j \in \mathcal{D}.
	\end{cases}
\end{equation}
Thus, $ \mathbf{f}$ can be obtained by solving the equation, which is experssed as $ 			\mathbf{f} = \sum_{k=1}^{K}\omega_k \mathbf{h}_k + \sum_{l=K+1}^{K+L}\omega_l \mathbf{h}_l$, where $ \omega_k = \sqrt{\frac{1-\sum_{l=K+1}^{K+L}\omega_l^2 \mathbf{h}_l^H\mathbf{h}_l}{(\mathbf{h}^H_k\mathbf{h}_k)^2\sum_{k=1}^{K}\frac{1}{\mathbf{h}^H_k\mathbf{h}_k}}} $ and $ \omega_l = \sqrt{\frac{E_0}{P_t(\mathbf{h}^H_l\mathbf{h}_l)^2}} $.

\end{appendices}

\bibliography{IEEEtran}
\bibliographystyle{ieeetr}

\begin{IEEEbiography}[{\includegraphics[width=1in,height=1.25in,clip,keepaspectratio]{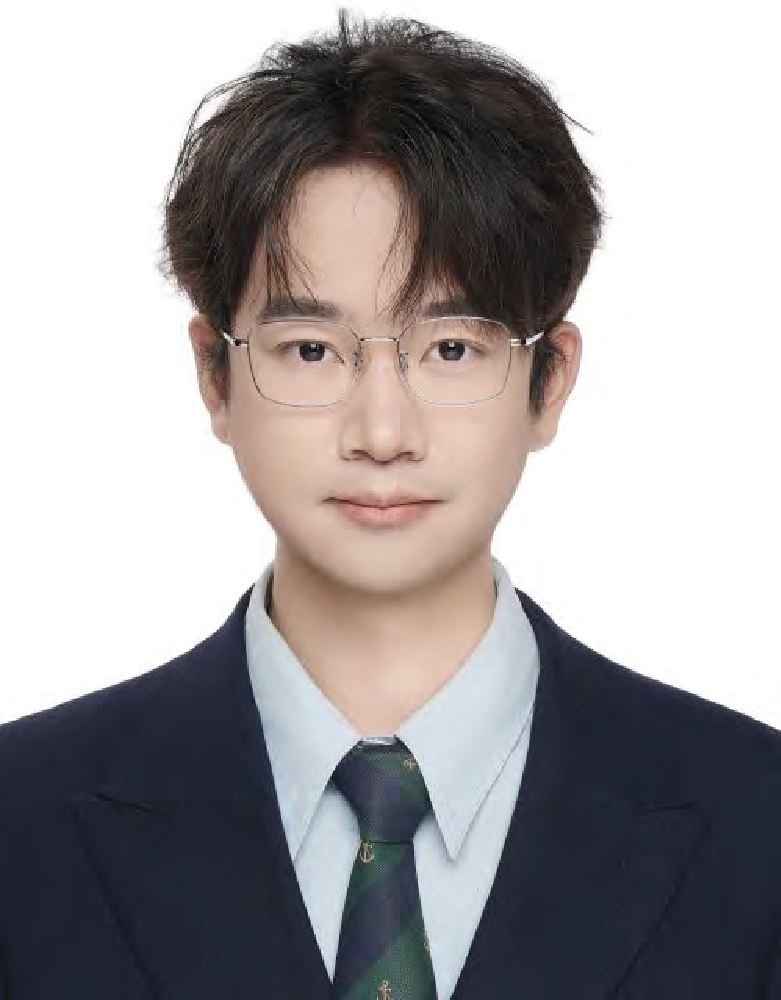}}]{Qingxiao Huang}[S'24]
	received his B.Eng. degree from Xidian Univesity, China, in 2022, and the M.Sc. degree from University of Electronic Science and Technology of China (UESTC), in 2025. He is currently pursuing the Ph.D. degree with the Department of Computer Science, City University of Hong Kong. His research interests include intelligent reflect surface networks, near-field communication and holographic MIMO. He has received several honors and awards, including the National Graduate Scholarship (2024), the Outstanding Master’s Thesis Award of UESTC and the Outstanding Graduate of Sichuan Province. He serves as a TPC Member for the IEEE VTC 2025-Fall Workshops.
\end{IEEEbiography}
\begin{IEEEbiography}[{\includegraphics[width=1in,height=1.25in,clip,keepaspectratio]{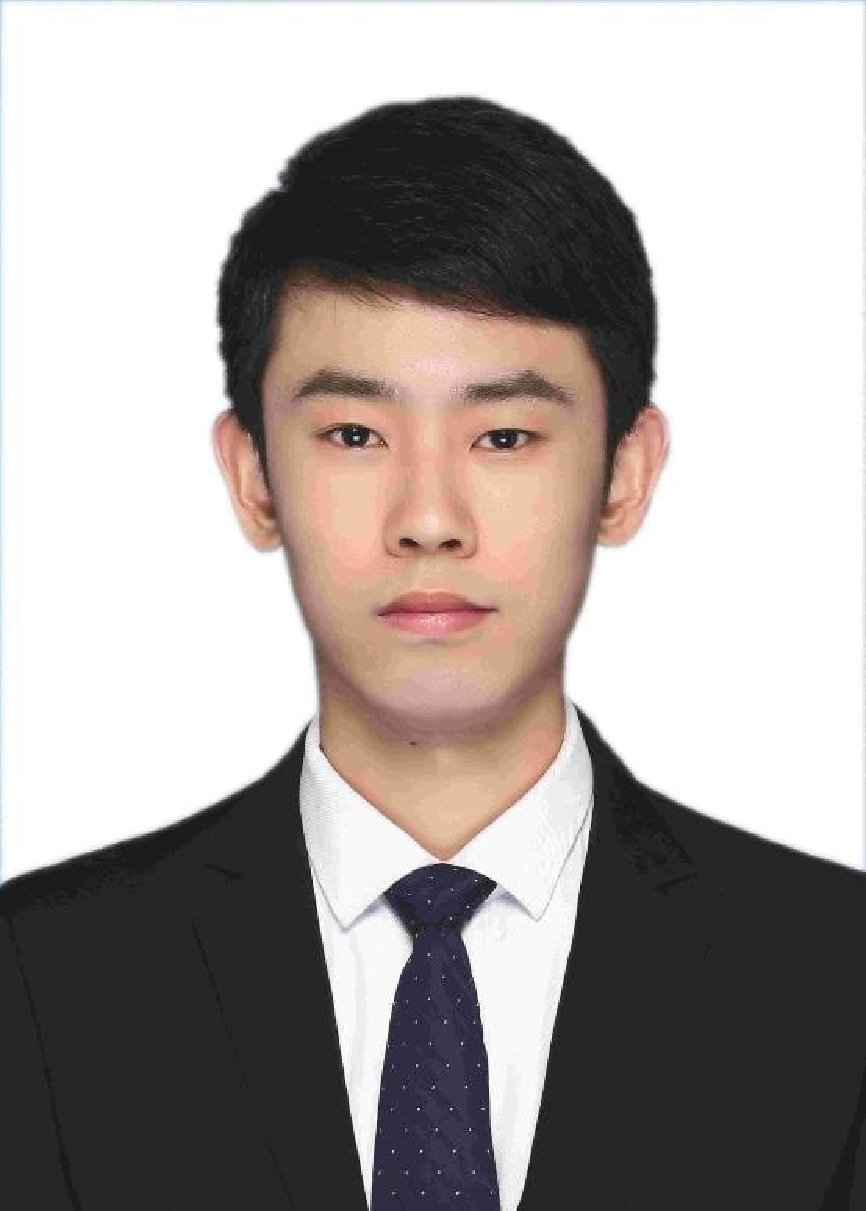}}]{Yizhe Zhao}[S'16, M'21] received the PhD in 2021 in School of Information and Communication Engineering from University of Electronic Science and Technology of China (UESTC), where he is currently an associate professor. He has been a visiting researcher with the Department of Electrical and Computer Engineering, University of California, Davis, USA. He is a member of IEEE and a senior member of China Institute of Communications. He is selected in Young Elite Scientists Sponsorship Program by China Association for Science and Technology (CAST). He serves for China Communications and Journal of Communications and Information Networks (JCIN) as the Guest Editor, and is also a TPC member of several prestigious IEEE conferences, such as IEEE ICC, Globecom. He was the recipient of IEEE CSE Best Paper Award in 2023. His research interests include modulation and coding design, integrated data and energy transfer, fluid antenna systems.
\end{IEEEbiography}
\begin{IEEEbiography}[{\includegraphics[width=1in,height=1.25in,clip,keepaspectratio]{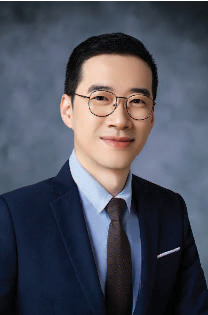}}]{Jie Hu}
	[S'11, M'16, SM'21] (hujie@uestc.edu.cn) received his B.Eng. and M.Sc. degrees from Beijing University of Posts and Telecommunications, China, in 2008 and 2011, respectively, and received the Ph.D. degree from the School of Electronics and Computer Science, University of Southampton, U.K., in 2015. Since March 2016, he has been working with the School of Information and Communication Engineering, University of Electronic Science and Technology of China (UESTC). He is now a Research Professor and PhD supervisor. He won UESTC's Academic Young Talent Award in 2019. Now he is supported by the ``100 Talents'' program of UESTC. He is an editor for \textit{IEEE Wireless Communications Letters}, \textit{IEEE/CIC China Communications} and \textit{IET Smart Cities}. He serves for \textit{IEEE Communications Magazine}, \textit{Frontiers in Communications and Networks} as well as \textit{ZTE communications} as a guest editor. He is a technical committee member of ZTE Technology. He is a program vice-chair for IEEE TrustCom 2020, a technical program committee (TPC) chair for IEEE UCET 2021 and a program vice-chair for UbiSec 2022. He also serves as a TPC member for several prestigious IEEE conferences, such as IEEE Globecom/ICC/WCSP and etc. He has won the best paper award of IEEE SustainCom 2020 and the best paper award of IEEE MMTC 2021. His current research focuses on wireless communications and resource management for B5G/6G, wireless information and power transfer as well as integrated communication, computing and sensing.
\end{IEEEbiography}
\begin{IEEEbiography}[{\includegraphics[width=1in,height=1.25in,clip,keepaspectratio]{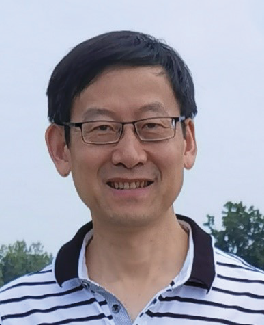}}]{Kun Yang}
	[F’22] received his PhD from the Department of Electronic \& Electrical Engineering of University College London (UCL), UK. He is currently a Chair Professor of Nanjing University and an affiliated professor at the University of Essex and UESTC. His main research interests include wireless networks and communications, communication-computing cooperation, and AI (artificial intelligence) for wireless. He has published 500+ papers and filed 50 patents. He serves on the editorial boards of a few IEEE journals (e.g., IEEE WCM, TVT, TNB). He is a Deputy Editor-in-Chief of IET Smart Cities Journal. He has been a Judge of GSMA GLOMO Award at World Mobile Congress – Barcelona since 2019. He was a Distinguished Lecturer of IEEE ComSoc, a Recipient of the 2024 IET Achievement Medals and the Recipient of 2024 IEEE CommSoft TC’s Technical Achievement Award. He is a Member of Academia Europaea (MAE), a Fellow of IEEE, a Fellow of IET and a Distinguished Member of ACM.
\end{IEEEbiography}
\begin{IEEEbiography}[{\includegraphics[width=1in,height=1.25in,clip,keepaspectratio]{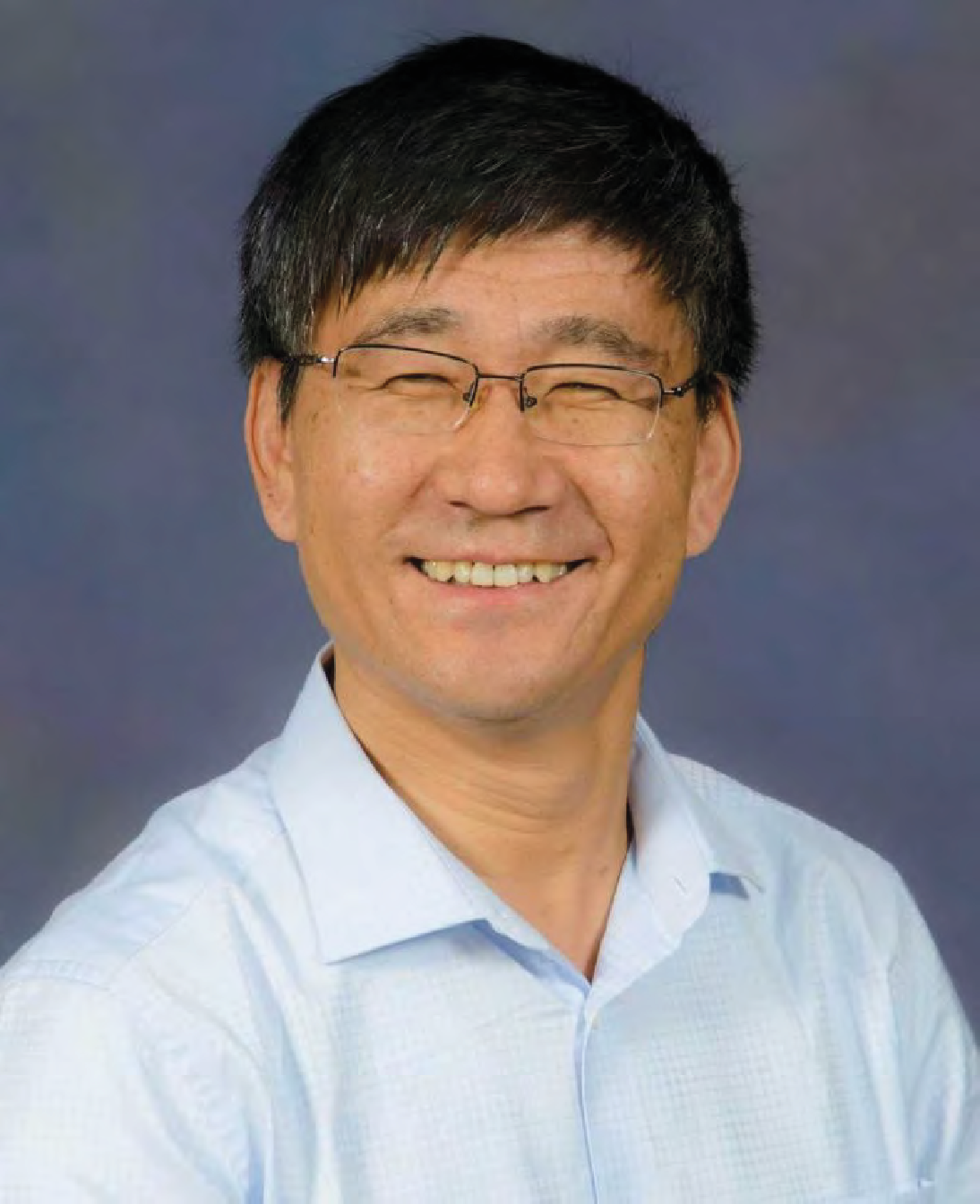}}]{Yuguang Fang}
	[S’92, M’97, SM’99, F’08] received an MS degree from Qufu Normal University, China, a PhD degree from Case Western Reserve University, USA, and another PhD degree from Boston University, USA, in 1987, 1994, and 1997, respectively. He joined the Department of Electrical and Computer Engineering at University of Florida in 2000 as an assistant professor, then was promoted to associate professor, full professor, and distinguished professor, in 2003, 2005, and 2019, respectively. Since 2022, he has been a Global STEM Scholar and Chair Professor with Department of Computer Science, City University of Hong Kong. He is the Founding Director of Hong Kong JC STEM Lab of Smart City funded by The Hong Kong Jockey Club Charities Trust. 
	
	He received many awards including US NSF CAREER Award, US ONR Young Investigator Award, the 2018 IEEE Vehicular Technology Outstanding Service Award, and several IEEE Communications Society awards (AHSN Technical Achievement Award, CISTC Technical Recognition Award, and WTC Recognition Award). He was the Editor-in-Chief of IEEE Transactions on Vehicular Technology and IEEE Wireless Communications. He is a fellow of ACM and AAAS. 
\end{IEEEbiography}
\end{document}